\documentstyle[12pt]{article}
\input{epsf}
\def \inbar{\vrule height1.5ex width.4pt depth0pt}
\def \C{\relax\hbox{\kern.25em$\inbar\kern-.3em{\rm C}$}}
\def \R{\relax{\rm I\kern-.18em R}}
\def \ids{ {\cal L}^{(2,\infty)}}
\def \idf{{\cal L}^{(1,\infty)}}

\def \idsf{{\cal L}^{(2,\infty)}({\cal H}_+,{\cal H}_-)}
\def \Gr{Gr_{1+}}
\def \Di{D_{1+}}
\newcommand{\Z}{\overline{Z}}
\newcommand{\beq}{\begin{equation}}
\newcommand{\eeq}{\end{equation}}
\newcommand{\bea}{\begin{eqnarray}}
\newcommand{\eea}{\end{eqnarray}}
\newcommand{\nn}{\nonumber}

\newcommand{\pdr}{\partial} 
 
\newcommand{\dt}{{\rm det}} 
\newcommand{\dth}{{\rm det}^{1 \over \hbar}} 
\newcommand{\eps}{\epsilon}

\newcommand{\Tr}{\hbox{Tr}}

\newcommand{\Hi}{{\cal H}}

\begin{document}

\title{Classical Mechanics and Geometric Quantization on 
an  Infinite Dimensional Disc and  Grassmannian}
\author{O.T.Turgut\\  
Institut Mittag-Leffler\\
   Aurav\"agen 17, S-182 62, Djursholm, Sweden\\
   and\\
   Department of Physics\\
 Bosphore University\\
  Bebek, Istanbul, Turkey\footnote{after March 99}}
\date{November  1998}
\maketitle
\begin{abstract}
 We discuss the classical mechanics on the Grassmannian and the 
Disc modeled on the ideal $\ids$.
We apply methods of geometric quantization to these systems.
Their relation to a flat symplectic space  is also 
discussed.

\end{abstract}

\section{ Introduction}

We will analyze geometric quantization of a classical system
which has as its phase space  the infinite dimensional
Grassmannian or  the Disc
modeled on the ideal $\idsf$.
There are two motivations for our work.
The classical dynamics studied  should correspond 
to the large-$N_c$ limit of a  quantum system which requires  a
logarithmic
renormalization.
Its quantization should give us an understanding of this system in 
the Schr\"odinger picture. This picture has some advantages 
over the scattering matrix, as well-known in the physics literature.
The second is to study and understand infinite dimensional systems,
their quantization should lead to some interesting mathematical
questions. A good example is typical two dimensional field theory models,
which do not require a renormalization but only a normal ordering
\cite{2dqhd,rajtur}.
It will be interesting to develop the necessary tools for more
complicated systems, and perhaps give a more precise meaning to
renormalized field theories.
We should add that, in this article we do not study any particular 
Hamiltonian, and associated delicate domain problems.
In some sense we have only made an attempt to 
study part of the kinematics.
The full understanding will require studying a specific model.

\section{The Disc and the  Grassmannian}

Our approach is inspired from  
the discussion of the Grassmannian in the book by Pressley and Segal 
\cite{presseg} and closely follows our previous 
work \cite{rajtur}.
We will extend some of our previous ideas to this case.

Let $\Hi$ be a separable infinite dimensional 
complex Hilbert space; $\Hi_-$ and $\Hi_+$ are two 
orthogonal isomorphic subspaces with $\Hi=\Hi_-\oplus \Hi_+$.  
Physically, one can think of these two spaces as the decomposition of the 
one particle Hilbert space into positive and negative energy states.

Define the Disc 
$D_{1+}(\Hi_-,\Hi_+)$ to be the set of 
all operators $Z:\Hi_+\to \Hi_-$ such that 
$1-Z^{\dag}Z>0$ and $Z$ is in $\idsf$. 
We refer to the Appendix for the ideals $\ids$ and $\idf$.

Since the space $\idsf$ is contractible, the set of $Z$ for the Disc
can be taken as a coordinate system.
It is an infinite dimensional complex manifold, modeled on a Banach space.

In a similar spirit,
we define the Grassmannian to be the set of
closed subspaces $W$  of $\Hi$, such that 
the projection ${\rm Pr}_+:W \to \Hi_+$ is Fredholm and
the projection ${\rm Pr}_-:W \to \Hi_-$ is in 
$\ids(W,\Hi_-)$.
We first define
natural group actions on these spaces.
These group actions are used to prove that these are
manifolds\footnote{generalizations of this kind has been 
pointed out in \cite{presseg}, and 
explored in \cite{mickraj} for 
the ideals ${\cal L}^p$. See also 
\cite{langmann}.}

We introduce  the following   pseudo-unitary group 
which is a subset of the invertible operators
from ${\cal H}$ to $\cal H$:
\beq
       U_{1+}({\cal H}_-,{\cal H}_+)
             =\{ g| g\epsilon g^{\dagger}=\epsilon, \quad g^{-1}\ {\rm
                    exists\ \ and } \ [\epsilon, g] \in \ids \}
.\eeq

Here $\eps=\pmatrix{-1&0\cr 0&1}$ with respect to the decomposition 
$\Hi=\Hi_-\oplus \Hi_+$.
If we decompose the matrix into block forms,
\beq
        g=\pmatrix{ a& b\cr c&d\cr}  \nn
\eeq
we have, $a:{\cal H}_- \to {\cal H}_-$, $b:{\cal H}_+ \to {\cal H}_-$
, $c:{\cal H}_- \to {\cal H}_+$ and $d:{\cal H}_+ \to {\cal H}_+$.  Then, the  
off-diagonal elements $b$ and $c$
 are in $\ids$  and the diagonal elements $a$ and $d$
are bounded operators. In fact they are
invertible  operators, since their spectrum does not contain zero.
The conditions on the off-diagonal elements imply some control over
how much $\Hi_+$ and $\Hi_-$ mix with each others.

We define an action of $U_{1+}({\cal H}_-,{\cal H}_+)$ on the Disc $\Di$:
\beq
     Z \mapsto g \circ Z=(aZ+b)(cZ+d)^{-1}.
\eeq
 The condition $1-Z^{\dag}Z>0$ implies that $cZ+d$ is invertible and bounded. 
Since the space of $\ids$  is a two-sided ideal, 
$(aZ+b)(cZ+d)^{-1}$  is still in $\ids$. Thus our  action is well--defined.
 
The stability 
subgroup of the point $Z=0$ is $ U({\cal H}_-) \times U({\cal H}_+)$, 
$U(\Hi_\pm)$
being the group  of {\it all} unitary  operators on $\Hi_\pm$.
Moreover, any point  $Z$ is the image of $0$ under the action of the group,
$g \circ (Z=0)=bd^{-1}$. (note that $bd^{-1}$ is in $\ids$ and
$d^{\dagger}d=1+b^{\dagger}b$ implies that $1-(bd^{-1})^{\dagger}bd^{-1}>0$).
We therefore see that $D_{1+}$ is a homogeneous space and
given by the quotient;
\beq
      D_{1+}= U_{1+}({\cal H}_-,{\cal H}_+) /
            U({\cal H}_-) \times  U({\cal H}_+)
.\eeq

It is possible to view $Gr_{1+}$ as a coset space of
complex Lie groups. Incidentally this will define  a complex structure on
$Gr_{1+}$ which will be useful for geometric quantization.
Define a  subset of the  general linear group
\beq
     GL_{1+}=\{\gamma|\gamma \ \hbox{\rm  is  invertible};
[\eps,\gamma]\in \ids\}.
\eeq
When we decompose g into $2\times 2$ submatrices
$\gamma_{12},\gamma_{21}\in \ids$ while $\gamma_{11}$ and $\gamma_{22}$
are Fredholm.
This is a Banach-Lie group modeled on 
$\idsf$.
We take the space of all endomorphisms on $\Hi$
with the same condition on the off-diagonals, ${\rm End}_{\ids}(\Hi)$,
and give it the natural topology
under the norm $||A||_+=||[\epsilon, A]_+||+||[\epsilon, A]||_{\ids}$.
The invertible elements of ${\rm End}_{\ids}(\Hi)$, is a group, open 
under  this topology, and it has a tangent space which comes from the
natural imbedding.
It is straightforward to define the ``Borel subgroup'';
\beq
     B_{1+}=\{\beta=\pmatrix{\beta_{11}&\beta_{12}\cr
0&\beta_{22}}|\beta\in GL_{1+}\}.
\eeq
This is the stability group of ${\cal H}_-$ under the action of
$GL_{1+}$ on ${\cal H}$. Thus the Grassmannian (which is the orbit
of ${\cal H}_-$) is the complex coset space,
\beq
     Gr_{1+}=GL_{1+}/B_{1+}.
\eeq

It will be convenient to use the following operators for the points on
 $D_{1+}$,
 $\Phi:{\cal H} \to {\cal H}$,
\beq
     \Phi=1-2\pmatrix{ (1-ZZ^{\dagger})^{-1}
                                 &-(1-ZZ^{\dagger})^{-1}Z\cr
                         Z^{\dagger} (1-ZZ^{\dagger})^{-1}
                      &-Z^{\dagger}(1-ZZ^{\dagger})^{-1}Z\cr}
.\eeq
One can see that under the transformation $Z \mapsto  g\circ Z$,
 $\Phi \mapsto g^{-1}\Phi g$. $\Phi$
satisfies $\epsilon \Phi^{\dagger} \epsilon=\Phi$ and
$\Phi^2=1$. Also, $\Phi-\epsilon \in \ids$, so 
that as an operator $\Phi$ does not differ from $\epsilon$ in an
 arbitrary way.

We can equivalently 
define the Grassmannian to be the  following  set of operators on $\cal H$:
\beq
     Gr_{1+}=\{\Phi|\Phi=\Phi^{\dag};\Phi^2=1;\ \Phi-\epsilon \in \ids \}.
\eeq

Since $\Phi^2=1$ and it is self adjoint, 
it can be diagonalized by the action of 
\beq
      U_{1+}({\cal H})=\{g|g^{\dag}g=1;[\eps,g]\in \ids\}.
\eeq
Let  us split $g$ into $2\times 2$ blocks
\beq
     g=\pmatrix{g_{11}&g_{12}\cr g_{21}&g_{22}}. \label{sp1}
\eeq
 The convergence condition on $[\eps,g]$ is the  statement
that the off--diagonal blocks $g_{12}$ and $g_{21}$ are in
$\ids$. It then follows, that $g_{11}$ and $g_{22}$ are
Fredholm operators. The Fredholm index of $g_{11}$ is opposite to
that of $g_{22}$; this integer is a homotopy invariant of
$g$ and we can decompose $U_{1+}(\cal H)$ into connected components
labeled by this integer.

We can see that  $U_{1+}(\Hi)$ is a real form of this group.
$GL_{1+}(\Hi)$ is the topological product of $U_{1+}(\Hi)$ and the
contractible space of positive definite elements by using the fact
that ${\rm End}_{\ids}(\Hi)$ has square-root of positive 
elements  well-defined and continuous
under its topology.\footnote{we can show that 
$U_{1+}$ is a deformation retract of $GL_{1+}$, similar 
to the finite dimensional case.} 

With the projection $g\to g\eps g^{\dag}$, we see that
$ Gr_{1+}$ is a homogeneous space of $U_{1+}(\cal H)$:
\beq
      Gr_{1+}= U_{1+}({\cal H})/U({\cal H}_-)\times  U({\cal H}_+).
\eeq
Any $\Phi\in Gr_{1+}$ can be diagonalized by an element of
$ U_{1+}({\cal H})$, $\Phi=g\eps g^{\dag}$; this $g$ is ambiguous up to
right multiplication by an element that commutes with
$\eps$. Such elements form the subgroup
\beq
      U({\cal H}_-)\times  U({\cal H}_+)=\{h|h=\pmatrix{h_{11}&0\cr
0&h_{22}}; h_{11}^{\dag}h_{11}=1=h_{22}^{\dag}h_{22}\}.
\eeq
Each point $\Phi\in  Gr_{1+}$ corresponds to a subspace of
$\cal H$: the eigenspace of $\Phi$ with eigenvalue $-1$. Thus
$Gr_{1+}$ consists of all subspaces obtained from ${\cal H}_-$ by an
action of $U_{1+}$.

To define the tangent space at each point we can use the 
action of the group on itself.
For our purposes it is better to take  the 
group action on the left.
Since the stability subgroup of $\epsilon$ is 
$U({\cal H}_-)\times  U({\cal H}_+)$ in both cases 
the tangent space is isomorphic to 
the corresponding off-diagonal algebras. 
In each case this is 
equivalent to $\idsf$ as a vector space, due to 
hermiticity(or psedo-hermitcity) condition. 

Any given $u \in U_{1+}$ defines a vector at a given point, 
and a vector field can be expanded in terms of the 
local set of vectors.
The action of a vector field on $\Phi$
is given by 
$V_{u(\Phi)}(\Phi)=[u(\Phi), \Phi]=g[g^{-1} u(\Phi)g,\epsilon]g^{-1}$.
The tangent space has a set of vectors which are given by the 
completion of the finite rank operators inside 
$\ids$. This is also an ideal inside ${\cal B}=B(\Hi)$
and is a separable Banach space under the same norm as 
$\ids$, we will denote this set by $(\ids)^{(0)}$.
The tangent space has a non-canonical decomposition at each point 
which is isomorphic to 
$(\ids)^{(0)} + \ids/ (\ids)^{(0)}$,
the second part is a ``transversal piece''.
As we will see this quotient will be important for the 
dynamical system we have in mind.

We introduce  the cotangent space as a 
formal expression
\footnote{
The dual space requires more care in infinite dimensions.
We can think of  
the norm dual of $\ids$, yet this space does not have
a simple characterization.
If we assume that the tangent space is in fact the 
norm dual of the cotangent space, we have 
a simple description of the cotangent space.
We refer to 
Gohberg and Krein for the details \cite{gohberg}. 
In this article  we will leave the question of the dual open,
and use one-forms only when we have an explicit formula.}
$d\Phi$ via its contraction with the vector field at a given point;
$d\Phi\big(V_{u(\Phi)}\big)=V_{u(\Phi)}(\Phi)$.

We would like to think of the $D_{1+}$ and $Gr_{1+}$ as classical phase
spaces.
To do this we need to introduce a Poisson bracket.
We will search for a symplectic form on this space.
It is tempting to generalize the finite dimensional formula to this
case.
If we write down 
$\Omega={i \over 4} \Tr \Phi d\Phi \wedge d\Phi$ we see that the trace
in general does not exist.
However one can see that the divergence is logarithmic, in fact the 
formal expression $\Phi d\Phi\wedge d\Phi$ belongs to 
$\idf$.
Hence we can replace the ordinary trace by the Dixmier trace.
Dixmier trace is used in non-commutative geometry, for a masterful
presentation of its properties we refer to the book and lecture notes
of Connes \cite{connes2}.

Each choice of the trace will give another symplectic form, 
they all agree  on the ``measurable'' part of the ideal 
$\idf$. Since the ``measurable'' elements do not form an ideal, we
cannot
assume that the symplectic form is independent of the choice of the
limit point.

Another important point is to remember that the 
Dixmier trace vanishes on the ideal 
generated by the completion of finite rank operators,
$(\idf)^{(0)}$. 
In the applications one  expects that the operators we have to
consider are pseudo-differential operators on manifolds. 
The physically relevant group of transformations 
are modeled on pseudo-differential operators 
which belong to the specific ideals 
that we have defined.
In the case of classical pseudo-differential 
operators, the Dixmier trace is uiquely defined;
it is  equal to the Wodzicki residue of the pseudo-differential operator,
as shown by Connes \cite{connes1}.
An interesting applications of Dixmier trace to the class of 
elliptic pseudodifferential operators is given in 
\cite{kontsevich}, and to chiral anomaly in \cite{mickellson}.
An interesting discussion of the central extensions and Schwinger terms are 
given in \cite{ferreti}.

In this article we will consider the general case, and show the 
dependence of the symplectic form to this limiting process $\omega$,
explicitly on our definition of the symplectic form:
\beq
     \Omega_\omega= {i\over 4}\Tr_\omega \Phi d\Phi\wedge d\Phi.
\eeq 
The existence of such a trace  is the reason for our choice $\ids$.
One can check that the above form is closed;
it is not so obvious that it is nondegenerate.
In fact it vanishes whenever the result of the contractions 
with the vectors at a given point is in   the completion of 
the finite rank operators
inside  $\ids$, $(\ids)^{(0)} \neq
\ids$.
This completion is a separable Banach space,
and an ideal inside ${\cal B}$ as well, 
whereas $\ids$  is a nonseperable Banach space.

The above form is invariant under the action of $U_{1+}(\cal H)$ for the
$Gr_{1+}$  and invariant under the action
 of $U_{1+}({\cal H}_-,{\cal H}_+)$
for the $D_{1+}$. The formal expression is defined as
\beq
  i_{V_u}i_{V_v}\Omega_\omega=
       {i \over 8}\Tr_\omega \Phi[[u,\Phi],[v,\Phi]]
\eeq
One can show the invariance using this expression immediately
(see below).
Thus,  $Gr_{1+}$  and $D_{1+}$ are both
homogeneous manifolds with an invariant closed two-form  similar to  the finite
dimensional case. 

Unfortunately this form is degenerate;
it vanishes on the part of the tangent space which 
corresponds to $(\ids)^{(0)}$\footnote{This means that we throw
 away a large part of
the symplectic manifold. Perhaps a physically more 
appropriate choice is to use a combination, which 
keeps the information about the ``small'' directions.
Although this can be done we will focus on the above 
symplectic form.}.
To see this let us calculate the contraction of $\Omega_\omega$ at a
point $\Phi$, with a vector field which belongs to the 
part $(\ids)^{(0)}$. Let us assume that this vector is generated by $u$
acting from the left. The condition for the vector to be in
$(\ids)^{(0)}$ is  simply $[\epsilon, g^{-1} u g] \in (\ids)^{(0)}$ at
the point $\Phi=g \epsilon g^{-1}$.
\beq
     i_{V_u}\Omega_\omega={i \over 8}\Tr_\omega \Phi [[u, \Phi],d\Phi]
\eeq
We will show that the contraction of this one form with an arbitrary 
vector on the tangent space at the same point $\Phi$
is zero, hence the form is zero.
Any such vector on the tangent is again generated by the left action
with a Lie algebra element $v$,
\beq
 i_{V_v}i_{V_u}\Omega_\omega={i \over 8}\Tr_\omega \epsilon [[\epsilon,
  g^{-1} ug],[\epsilon, g^{-1} v g]]
\eeq
Using $[\epsilon, \pmatrix{{\cal B} &\ids \cr \ids &{\cal
B}}]=\pmatrix{0 &\ids \cr \ids &0}$
and similarly for the other part, we have,
\beq
   \pmatrix{0 &\ids \cr \ids &0}\pmatrix{0 &(\ids)^{(0)} \cr
(\ids)^{(0)} &0}=\pmatrix{(\idf)^{(0)} &0 \cr 0 &(\idf)^{(0)}}
,\eeq
where we use $(\ids)^{(0)}\ids \in (\idf)^{(0)}$(see Appendix 
for a proof).
The Dixmier trace vanishes on $(\idf)^{(0)}$ and this shows that 
the form is zero.
In general since the tangent space  has the direction given by 
$(\ids)^{(0)}$, the cotangent space has one-forms 
which do not vanish on them.
If $\Lambda$ is a form such that $\Lambda(V_u)\neq 0$,
then $\Lambda=i_Y \Omega_\omega$ has  no solution for the vector  $Y$.
If we assume that the de Rham theory makes sense on these spaces,
since $\Di$ is contractible, and $\pi_1(\Gr)=0$, we would expect that
there is a function 
$f$ such that 
$\Lambda=df$, and this will   show that one cannot obtain 
Hamiltonian vector field for any given
function $f$ in general.
Nevertheless, as we will see,  for the relevant  part of the 
space, that is for ``large'' motions, directions which 
belong to $\ids /(\ids)^{(0)}$, the form is non-degenerate.
This will allow us to define classical dynamics for certain 
systems.

Before we continue, let us  point out an  important  observation.
For clarity let us concentrate on $\Gr$.
There is an interesting leaf of $\Gr$ which corresponds to the
orbit of $\epsilon$ under the following subgroup;
\beq
  U_{1+}^{(0)}=\pmatrix{{\cal B}&(\ids)^{(0)} \cr (\ids)^{(0)} &{\cal B}}
\eeq
We denote this orbit by $\Gr^{(0)}$.
Since the connected components are still labeled by the integers, 
each connected component of $\Gr$ has the 
connected component of $\Gr^{(0)}$ inside.
If we take $\Gr^{(0)}$'s  own tangent space, generated by 
$U^{(0)}_{1+}$,  the symplectic form vanishes on this 
orbit using the fact that $(\ids)^{(0)}\ids \in (\idf)^{(0)}$,(see
Appendix for a proof).
The same remarks apply to the Disc $\Di$, and we spare the details for
the reader.

Since the group action preserves the two from $\Omega_\omega$ 
the Lie derivative along the direction of any vector field generated
by the group action gives us zero.
This raises the possibility of finding the moment 
maps which would generate 
the infinitesimal action of
$U_{1+}({\cal H}_-,{\cal H}_+)$ and $U_{1+}({\cal H})$ respectively.
We can start with the finite dimensional answer;
one can check that to avoid the divergence   
the finite dimensional answer has to be modified as
$-\Tr_\omega^\epsilon  u(\Phi-\epsilon)$, where $u$ 
is a hermitian matrix which is in the Lie algebra of $U_+(\Hi)$ for the
$Gr_{1+}$ and a pseudo hermitian ($u^{\dag}=\eps u\eps$) operator
which 
belongs to $U_{1+}(\Hi_-,\Hi_-)$ for $D_{1+}$. 
We use a conditionally convergent trace for the 
variable $\Phi-\epsilon$ if we think of the group acting from the
left.
To see that the above expression makes sense 
consider  $u \in\pmatrix{{\cal B}&\ids \cr \ids &{\cal B}}$,
and $\Phi-\epsilon \in\pmatrix{\idf &\ids\cr \ids &\idf}$. (Here
 ${\cal B}$ is the space of bounded operators.)
As a result $u(\Phi-\epsilon) \in \pmatrix{\idf &\ids\cr \ids &\idf}$.
Let us define $M=\Phi-\epsilon$.
Formally  the equation for the 
group action is satisfied,
\beq
      -\Tr^\epsilon_\omega u d\Phi=\Omega_\omega(V_u,\   ) \ \ \
     V_u(\Phi)=[u, \Phi]
\eeq
but we must be careful since if $g^{-1} ug \in (\ids)^{(0)}$ the 
right handside vanishes as we have shown.
The only way to satisfy this equation, is to show that whenever 
the right handside vanishes the left vanishes as well.
This can be checked as follows;
\beq
  df_u(V_v)=\Tr^\epsilon_\omega u[v,\Phi]=\Tr_\omega^\epsilon g^{-1} u
g[\epsilon, g^{-1} v g]
,\eeq
for any $v \in U_{1+}$.
By the same argument as above, we see that the resulting expression
inside the trace is of class $(\idf)^{(0)}$.
Hence the trace is also zero.

We notice  that the form corresponding to the
moment map $-\Tr^\epsilon_\omega(uM^{(0)})$ for any
$u \in U_{1+}$ when contracted with the elements of the tangent space 
of $\Di^{(0)}$ or  $\Gr^{(0)}$ also gives zero.
This is in some sense the orbit one can 
neglect; we can think of it as the null orbit. 
This shows that we cannot take the 
separable part in our definition of the group 
$U_{1+}$, denoted as 
$(U_{1+})^{(0)}$,  if we use the Dixmier trace.
In a recent preprint \cite{dykema}, it is argued that there  is no
positive trace on the ideal $(\ids)^{(0)}$, this implies that a similar 
construction cannot be achived for the group $(U_{1+})^{(0)}$.
 
However, it is not enough to show that the 
moment functions satisfy a 
consistent equation. Normally in the Hamiltonian
formalism we are given a 
function and asked to find the 
vector field generated by this function.
Our discussion shows that this vector field at every point is only
determined in the eqivalence class $\ids / (\ids)^{(0)}$.
 It is easy to see 
that if we take instead of $u$, a vector generated 
by $u+v$ such that $g^{-1}vg \in (\ids)^{(0)}$, 
$-df_u={i \over 8}\Tr_\omega\Phi[[u+v,\Phi],d\Phi]={i\over
8}\Tr_\omega
\Phi[[u,\Phi],d\Phi]+{i \over 8}\Tr_\omega[[v,\Phi],d\Phi]$ for  the 
last piece is zero by the previous argument. 
This implies that the moment funtions do not have unique vector
fields,
they  generate the motions in the
``transversal direction'' $\ids / (\ids)^{(0)}$ with an 
undetermined   piece in $(\ids)^{(0)}$.
Let us look at the infinitesimal part of this 
evolution for two different choices of the vector field in the
equivalence class;
$\Delta \Phi=t[u, \Phi]$ and $\Delta' \Phi=t[u',\Phi]$.
The difference of these two infinitesimal evolutions 
are given by 
$\Delta\Phi-\Delta'\Phi=[u-u',\Phi]=g[g^{-1}(u-u')g,\epsilon]g^{-1}$.
Since the ambiguity is a result of  the difference,
which satisfies $[g^{-1}(u-u')g,\epsilon] \in (\ids)^{(0)} $; 
this term is in the orbit  $\Di^{(0)}$, or in  the 
same connected component of $\Gr^{(0)}$.
Thus the difference of the infinitesimal evolutions can be conjugated
to the orbit $\Di^{(0)}$, or to the same connected component of
$\Gr^{(0)}$.
This implies that the  relevant space for the 
classical dynamics is not the original quotient we look, but a 
smaller one, given by 
$U_{1+}/U_{1+}^{(0)}$.
In fact as we will see later on, this reduction has an interesting 
consequence.
But, for the moment we will continue to use  the ``unreduced phase
space''.

To talk about classical evolution, we 
will make the proposal that this type of classical systems 
are defined through an equivalence relation.
We will assume that two dynamical evolutions are equivalent if they
could be conjugated to the ``null'' orbit, $\Di^{(0)}$, for 
$\Di$ and to the same connected component of $\Gr^{(0)}$, for 
$\Gr$.
Another way to think about this is that the dynamics in the 
``small'' directions cannot be determined. 
The moment functions we consider only determine the evolution 
under equivalence.

This feature will affect the dynamics generated by Hamiltonians 
of more complicated functions.
The generic Hamiltonians we have in mind are quadratic functions of
the variable $M=\Phi-\epsilon$ plus a moment map.
We think of the moment map as the free part of the Hamiltonian
since it generates 
the goup action upto an equivalence, and the 
quadratic piece as an ``interaction''.

Formally we can write the Hamiltonians as,
\beq 
 h=i\Tr_\omega^\epsilon uM+\Tr_\omega^\epsilon \hat K(M)M
\eeq
where, $\hat K$ is a linear  operator which acts on 
the variable $M$.
If we specify a basis, it can be expressed as
$(\hat K(M)M)^k_p=\sum_{ijk}K^{ik}_{jl}M^j_iM^l_p$.
The summability properties of the kernel $K^{ik}_{jl}$ should be 
such that the resulting 
operator is in the ideal 
$\idf$. For example, this can be achieved, if the map
$\hat K$ produces an operator in 
the group $U_{1+}$.  
Notice that it is important to use the Dixmier trace again.
As we mentioned in the moment maps, since the symplectic form
$\Omega_\omega$ vanishes upon contraction with
elements of the tangent space in the ``small'' directions, the 
interaction hamiltonian must have the same property.
As one can check 
this form of the Hamiltonian, when differentiated, gives 
a form which vanishes on the same subspace.
This will impose certain conditions on the choice of kernels.
Interactions which are ``too weak'' will not affect the 
equations of motion.

Let us give a typical interaction Hamiltonian.
We can take two moment functions and take their products;
\beq
    \sum_{ij}K^{ij} \Tr_\omega^\epsilon u_iM \Tr_\omega^\epsilon u_j M
\eeq
If the sum is over a finite number of terms then clearly 
this  is a well-defined expression.
In general one should be able to choose $K^{ij}$ such that the 
above expression is finite.
For these Hamiltonians, one can directly calculate 
the equations of motion by using the 
Poisson bracket relations among the 
moment functions only.
As we will see later on, they also have a 
simpler description in the quantum case.

A word of caution should be said here. Typically, the Hamiltonians 
are more singular than the symplectic form, and they require further
 renormalizations. This implies a choice of domain for the
Hamiltonian.
This should restrict the accessible regions of the phase space, or the
true phase space of the theory, and it may change  
the formulation of the problem drastically.
In real physical systems, we expect these problems to modify the 
precise formulation of the field theory. In this work, we do not
discuss
this more difficult problem.
In some sense this is part of the  kinematics, although interactions 
may even change this. 

There will always be vector fields generating the equations of motion 
upto equivalence.
In infinite dimensions determination of the 
integral curves and  their completeness  
are highly  non-trivial issues; the
answer depends on the Hamiltonian as well.
This is the classical version of the unitarity 
condition in quantum mechanics.

From general principles we expect 
that the Poisson brackets of the moment functions will  provide 
a realization of the corresponding Lie algebra possibly with a central
extension. We can calculate the Poisson bracket of two moment 
functions using a formal manipulation  and it gives us,
\beq
     \{f_u,f_v\}=f_{-i[u,v]}-i\Tr_\omega^\epsilon [\epsilon, u]v
.\label{moment}
\eeq
This is  
 a central extension of the full group, indexed by the
choice of limit process $\omega$. (Note that there is no 
ambiguity in this relation since the right handside 
is the same for equivalent choices of 
the vector fields).
One can explicitly check that 
$c_\omega(u,v)=\Tr_\omega^\epsilon [\epsilon,u]v$ satisfies
$c_\omega(u,v)=-c_\omega(v,u)$ and 
the cocycle condition,
$c_\omega(u,[v,w])+c_\omega(v,[w,u])+c_\omega(w,[u,v])=0$.
Since the 
 central term vanishes on the ideal $(\idf)^{(0)}$ it will 
not be there  when $u$ or $v$ $\in U_{1+}^{(0)}$.
The actual computation should be done with care, 
due to infinite dimensionality.
One can see that the both sides of the 
equation after explicit calculation give the same result, 
hence they are identical.
We will leave the details to the reader.

This gives us a symplectic realization of the Lie algebra of 
$U_{1+}(\Hi)$ for $\Gr$ and $U_{1+}(\Hi_-,\Hi_+)$ for $\Di$
except a central term.
Since the calculation of the both sides 
are actually zero whenever 
$g^{-1} u g$ or $g^{-1} v g  \in  U_{1+}^{(0)}$, this 
expression should be thought of as a realization of the 
``large'' part of the Lie algebra. 
\footnote{It is  more natural to look into  the spaces 
which are modeled on the quotients $\ids / (\ids)^{(0)}$.
Since $(\ids)^{(0)}$ is closed the 
quotient is well-defined.  Later on we will consider this 
point of view.} 
Let us consider this central extension for the 
pseudodifferential operators\footnote{This is pointed out to me
by Mickelsson};
then we have $\Tr_\omega[\epsilon,u]v={\rm Res}([\epsilon,u]v)$.
We recall that the residue is actually defined for all pseudodifferential
operators and it satisfies Res$[A,B]=0$\cite{wodzicki,fedos}.
This implies that, in fact 
 $\Tr_\omega[\epsilon, u]v=\pdr\phi(u,v)$, for  $\phi(u)=$Res$(\epsilon u)$,
hence 
the central extension is trivial.

Because of the infinite dimensionality an attempt to remove the
central term, in the genral case,  will result in a divergent expression.
This gives us   the Lie algebra of the non--trivial central extension of
$U_{1+}$ corresponding to $\omega$. 
We expect that these extensions are not equivalent in general for
different choices of 
$\omega$ but they all agree on the  subset  modelled on the 
``measurable'' elements--this is not a subalgebra of 
$\underline{U_{1+}}$, so one cannot 
reduce it to  this case.

\section{Quantization}

We continue to think of classical mechanics in geometric
terms\cite{arnold,abraham}. 
Let us
assume that
the phase space,
$\Gamma$ is a smooth manifold. 
If we have a Poisson structure on 
 the algebra of smooth functions $C^\infty(\Gamma)$, 
we can introduce classical dynamics.
Quantization of this classical system is 
given by a representation of the Poisson algebra of smooth functions 
by self-adjoint operators on a Hilbert space.
This is an overambitious program; in general 
there is no way to find such a representation.
The difficulties and various  methods have been well-explained in the
literature\cite{axelrod, hurt, krillov}.
In this article we will follow our point of view in 
\cite{rajtur}. We will find a represention of the 
Poisson algebra of moment functions.
Any composite function, which is related to the 
product of two moment maps can be qantized by giving an ordering rule.
We will not attempt to establish this idea in the present article.

Before we proceed further, we need to make a digression and introduce
a generalized ``determinant''.
It does not satisfy all the properties of a determinant.
As we will see, to think of it as a determinant simplifies 
the calculations.
Let us define ${\rm det}_\omega(1+A)$ for $A \in \idf$ as
\footnote{One  way to motivate  this definition is the
following.
Let us take the determiant formula, 
$\log {\rm det}(1+A)=\Tr\log(1+A)$.
We replace the trace by the Dixmier trace,
and define
$\log \dt_\omega (1+A)=:\Tr_\omega\log(1+A)=\Tr_\omega(A-{1 \over 2}A^2
+{1 \over 3}A^3-...)=\Tr_\omega(A)$ since all the higher terms 
are in ${\cal L}^1$ and the Dixmier trace vanishes on them.
Ths gives the above formula again.}

\beq
    {\rm det}_\omega(1+A)=\exp(\Tr_\omega(A))
\eeq
One can see that it satisfies the multiplicative property of the
determinants;
\beq 
   {\rm det}_\omega\big((1+A)(1+B)\big)={\rm det}_\omega(1+A)
   {\rm det}_\omega(1+B) 
,\eeq
due to $\idf \idf \in {\cal L}^1$ and $\Tr_\omega$ vanishes on
the trace class operators ${\cal L}^1$.
An interesting  property is that $\dt_\omega$  never vanishes.

We will 
use this to provide a representation of
the borel subroup on $\bf C$,  and  attempt to 
follow the geometric quantization program.

We will introduce an ad hoc representation, which comes from the 
geometric quantization performed in \cite{rajtur}, of the 
Lie algebra of the group $U_{1+}(\Hi_-,\Hi_+)$ on the space 
of ``holomorphic functions''\footnote{The author is not aware of
a well-established definition of 
holomorphicity for infinite dimensions. We assume 
that the algebraic operations on the coordinate $Z$
after the application of a dual element, if it is finite, provides 
a holomorphic function. This will be used for the grassmanian as well.}
on $\Di$.
\beq
   \hat f_{-u}\Psi(Z)=-i\hbar[{\cal L}_{V_u}\Psi(Z) -{1 \over \hbar}
       \Tr_\omega (\gamma Z)\Psi(Z)]
,\eeq
where $u \in {\underline U}_1({\cal H}_-,{\cal H}_+)$,
 and $V_u=V^Z_u\partial_Z+V^{Z^{\dagger}}
_u\partial_{Z^{\dagger}}$ is the formal vector field generated
by the action of $-u$.
It is easier to define the Lie derivative directly, by using the action
of the Lie algebra on the Disc.
We define this as,
\beq
{\cal L}_{V_u}\Psi(Z)=\lim_{t \to 0}
 {\Psi(Z +t(\alpha Z+\beta-Z\gamma Z-Z\delta))-\Psi(Z) \over t}
,\eeq 
where,
\beq
       u=\pmatrix{\alpha & \beta\cr
                  \gamma & \delta\cr}
\nn\eeq
is the decomposition of $u$ into block form; $\alpha^{\dagger}=-\alpha$
, $\beta^{\dagger}=\gamma$ and $\delta^{\dagger}=-\delta$ and further
$\gamma,\ \beta \in \ids$.
We notice that this differs from the finite dimensional answer by a 
constant term, which is infinite in this case.
The changes introduced in the functions are all 
``holomorphic'', so the action of the operators corresponding  to the
moment functions preserve the 
``holomorphicity condition''. 
 In order to show that this is the correct representation, we need 
to prove that the 
commutation relations are satisfied acting on these set of functions.

let us check that the commutation relations will give us a realization
of the Poisson bracket relations satisfied by the 
moment maps.
\beq
 [\hat f_{-u_1}, \hat f_{-u_2}]\Psi(Z)=
  i\hbar \hat f_{[u_1, u_2]} \Psi(Z)+i\hbar 
\Tr^\epsilon_\omega([\epsilon, u_1]u_2)\Psi(Z).
\eeq
A calculation shows that,
\beq
 [\hat f_{-u_1}, \hat f_{-u_2}]\Psi(Z)=
 \hbar^2\big({\cal L}_{u_1}{\cal L}_{u_2}-{\cal L}_{u_2}{\cal
L}_{u_1}\big)\Psi(Z)
+\hbar[ ({\cal L}_{u_1}\Tr_\omega \gamma_2 Z)-
({\cal L}_{u_2}\Tr_\omega \gamma_1 Z))]\Psi(Z) \nn
.\eeq
For simplicity we use ${\cal L}_u$ instead of 
${\cal L}_{V_u}$.
This is equal to 
\beq
 [\hat f_{-u_1}, \hat f_{-u_2}]\Psi(Z)=\hbar^2{\cal L}_{[u_1,
u_2]}\Psi(Z)+\hbar \Tr_\omega \{\gamma([u_1, u_2])Z\}
\Psi(Z) +\hbar \Tr_\omega(\gamma_1\beta_2-\gamma_2\beta_1)\Psi(Z)
\eeq
The last term, which is a constant multiple,  can be rewritten as 
$\Tr^\epsilon_\omega[\epsilon, u_1]u_2$.
Hence we see that it is a representation of the 
Poisson bracket relations (\ref{moment}).

This representation can in fact be integrated to a representation of the 
group action on the space of holomorphic functions:
\beq
     \rho_\omega (g^{-1})\Psi={\rm det}_\omega^{-{1 \over \hbar}}(d^{-1}cZ+1)
                \Psi\big((aZ+b)(cZ+d)^{-1}\big)
.\eeq
These representations are labelled by $\omega$, the choice of limit point,
 and 
$\hbar$.  Since the 
determinant never vanishes and actually given by an exponential
$\hbar$  is 
any real number.
\footnote{This point of view on quantization for finite dimensional 
homogeneous spaces appeared in \cite{berezin}}
To justify this, we will compute the infinitesimal form of the 
representation and show that it is  given by the operators corresponding
to the moment maps.
We write explicitly,
\beq
\rho(g^{-1})\Psi(Z)={\rm e}^{-{1 \over \hbar} {\rm Tr}_\omega(d^{-1}cZ)}\Psi(g
 \circ Z)
\eeq
and evaluate,
\beq
 \lim_{t \to 0}\Big\{ {\rho_\omega(1+tu)-\rho_\omega(1) \over t}\Big\}\Psi(Z)=
  \hbar[{\cal L}_{V_u}-{1 \over \hbar}\Tr_\omega(\gamma Z)]\Psi(Z).
\eeq
So we see that the infinitesimal form is given by the moment map
operators.
We still have to check that this is a representation;
\beq
\rho_\omega(g_1)\rho_\omega(g_2)\Psi(Z)=c_\omega(g_1,g_2)
\rho_\omega(g_1g_2)\Psi(Z)
\eeq
where $c_\omega(g_1,g_2)$ is a central term, which satisfies,
$c_\omega(g_1g_2,g_3)c_\omega(g_1,g_2)=c_\omega(g_1,g_2g_3)c_\omega(g_2,g_3)$. 
Since the Disc is topologically trivial the central extension 
can be described by specifying a function from 
the cartesian product of the group to ${\bf C}$.
An explicit calculation which is given in the Appendix, shows that 
the group property is satisfied with a central term,
\beq
     c_\omega(g_1,g_2)={\rm det}_\omega^{1 \over \hbar}
[(d_1d_2)^{-1}c_1b_2+1]=
{\rm exp}({1 \over \hbar}\Tr_\omega [(d_1d_2)^{-1}c_1b_2])
.\eeq
But there is still one more point we need to consider.
Recall that the orbit of $\epsilon$ corresponding to the subgroup
$U^{(0)}_{1+}$, $\Di^{(0)}$,  when considered as a 
submanifold with its own tangent
space has no dynamics under our choice of the symplectic form.
Hence the Poisson algebra of moment maps restricted on this
submanifold is trivially true, being zero on both sides.
Whereas the representation space we choose, when restricted 
to the subspace $\Di^{(0)}$ provide a nontrivial representation
of the Poisson algebra of moment maps.
This can be rectified by selecting a subspace 
of holomorphic functions which remain constant 
on the orbit $\Di^{(0)}$.
As a result, we define the Quantum Hilbert space
to be 
\beq
 \Hi_Q=\{ \Psi(Z)| \Psi(Z) \ \ {\rm  holomorphic \ on }\ \Di \ \ {\rm and}\
 \ \Psi\big|_{\Di^{(0)}}={\rm constant}\}
\eeq
This condition is consistent with the assumption  that the dynamics along the
directions $(\ids)^{(0)}$ is unimportant.
If we consider $\epsilon$ to be the ``vacuum'' configuration, its
orbit under $U_{1+}^{(0)}$ is equivalent to this ``vacuum''
configuration.

For this choice of the Quantum Hilbert space, we will exhibit a
class of wave functions.
We are unable to prove that these are the only 
possible ones.
Since the wave functions should be constant on the 
orbit $\Di^{(0)}$,
it suggests the use of the Dixmier trace again.

We can compose polynomials in the variable $Z$.
The interesting thing is to note that we can go upto quadratic terms
only inside the Dixmier trace.
For any choice of $A_i \in \ids$, and any two
$B_j, B_k \in B(\Hi)$, we can form,
\beq
  \Tr_\omega(A_iZ) \quad\&\quad \Tr_\omega(B_jZB_kZ)
\eeq

Using the generalized H\"older inequality \cite{simon},
 we have the inequalities,
\beq
 |\Tr_\omega(A_iZ)| \leq ||A_i||_{\ids}||Z||_{\ids} \quad
{\rm and}\quad |\Tr_\omega(B_kZB_jZ)| \leq
 ||B_k||||B_j||||Z||_{\ids}^2
\eeq
These show the continuity in $Z$ and  
with respect to $A_i$ and  $B_j, B_k$.
We can compose products of these kind of functions; 
\beq
   \Psi(Z)=\prod_{i,j,k} \Tr_\omega(B_j ZB_kZ)\Tr_\omega(A_i Z)
\eeq
The reader can verify that any higher power of
$Z$ is irrelevant, so these are the 
only combinations we can make. Various superpositions of these
 functions will give us the set of wave functions. 
\footnote{One should perhaps compare this with the analysis given in 
\cite{rajtur} There, the holomorphic wave functions can be constructed
using the analogy with the finite dimensional case.
We can use the fact that the dual of ${\cal L}^2$ is itself and 
the dual of ${\cal L}^1$ is $B(\Hi)$.
We can write down 
a general holomorphic function as sums and products of the expressions
of the form,
$\Tr(A Z)$,$\Tr(B_1ZB_2Z...B_mZ)$, where $A \in {\cal L}^2$ and 
$B_k \in B(\Hi)$ for all $k=1,..m$.
Not all of them are linearly independent of course.
With the appropriate inner product the Quantum Hilbert 
space constructed out of these functions is isomorphic to the usual 
Fock space and it is a seperable Hilbert space.}

We also would like to point out an interesting property of our wave 
functions. Since we have obtained them through the Dixmier trace, 
 we effectively perform
``logarithmic wave function renormalization''.

The Quantum Hilbert space may be 
very large, or maybe very small, it depends on the size of the 
space of  
``holomorphic functions'' on $\Di$, and the inner product.
It is not clear how one should introduce a measure to define 
an inner product in this Quantum Hilbert space. 
The term Hilbert space is only justified by thinking of this
as a quantization of a classical system.
It should be possible to extend the work in \cite{pickrel}
to this case. Once this is done, the completion of the
above set of wave functions with respect to this measure, will 
be the quantum Hilbert space, $\Hi_Q$.

Next we will construct the quantum operators for 
the $\Gr$.
Our approach will be somewhat ad hoc again.
Since $\Gr$ is topologically nontrivial, it may not be  possible to
represent wave functions as functions on the grassmannian.
It is natural to  introduce them as sections of a complex line bundle.
However, we will see that there are non-constant holomorphic
functions on this grassmanian.\footnote{This can be contrasted with 
the \cite{presseg, rajtur}, there, the finite dimensional grassmanians
are dense inside the full grassmanian, and it is well-known that on a
compact complex manifold there are no non-constant holomorphic
functions.
In our case, completion of finite rank objects will not be equal to
the full space.}
For the quantization of our classical
system, we actually need the sections of a line bundle.
For this it is better to think of $\Gr$ as a qoutient 
of another pair of groups as in the case discussed by \cite{presseg, rajtur}.
It is possible  that the extension is nontrivial both topologically
and algebraically. 
 Essentially, we will enlarge  `numerator'
and `denominator' by the same amount, this will keep the ratio same.
We define the group
$\tilde G_{1+}$:
\beq
     \tilde G_{1+}=\{ (\gamma, q)| q \in GL({\cal H}_-) ;\quad \gamma \in
GL_{1+}({\cal H}) ,\quad \gamma_{11}q^{-1}-1 \in \idf \}
.\eeq
Here, $\gamma_{11}$ denotes the mapping
$\gamma_{11} :{\cal H}_- \to {\cal H}_-$ in the block form of the matrix
$\gamma \in GL_{1+}({\cal H})$.
One can prove that the set of $q$'s which satisfy this condition is 
not empty using the definition of the 
group $GL_{1+}$ following Pressley and
Segal\cite{presseg}.
We can give a topology to this space using the two
topologies inherited from the 
bounded and $\idf$.
(Notice that 
the extension for any two sided ideal is mentioned in
\cite{presseg} at page 98.)

$\tilde G_{1+}$ is a complex Banach-Lie group under the multiplication
$(\gamma, q)(\gamma',q')=(\gamma\gamma',qq')$.
 We introduce $\tilde B_{1+}$, a closed complex subgroup of $\tilde G_{1+}$;
\beq
   \tilde B_{1+}=\{ (\beta, t)| \beta \in B_{1+}, t \in GL({\cal H}_-),
                \beta_{11}t^{-1}-1 \in \idf \}
\eeq
There is an action of $\tilde B_{1+}$ on $\tilde G_{1+}$. Since this
action  does not involve anything but multiplication in the group, it
is
holomorphic.
We enlarged $GL_{1+}({\cal H})$ and  $B_{1+}$ with the same set of elements, thus the
quotient is still the same;
\beq
    \tilde B_{1+} \to \tilde G_{1+} \to \Gr
.\eeq
In this case as well,  there are  
subgroups corresponding to 
completions of the finite rank elements.
These subgroups now can be written as, 
\beq
     \tilde G_{1+}^{(0)}=\{ (\gamma^{(0)}, q)| q \in GL({\cal H}_-)
;\quad \gamma^{(0)}
 \in
GL_{1+}^{(0)}({\cal H}) ,\quad \gamma_{11}^{(0)}q^{-1}-1 \in (\idf)^{(0)} \}
.\eeq
Existence can be proved along the same lines.
The stability subgroup also changes to $\tilde B_{1+}^{(0)}$, which is
defined similarly.
One can see that 
$\tilde G_{1+}^{(0)} \subset \tilde G_{1+}$ as a closed 
subset.
This subgroup will correspond to the null orbit, and it will not have
any extension.
 
Now, we can introduce the holomorphic line bundle corresponding to the
representation
$\rho(\beta, r)=\dt_\omega^{1 \over \hbar}(\beta_{11}r^{-1})$. 
There is no condition for the number 
$\hbar$ to be an integer  at this stage, since 
our definition for $\dt_\omega$ has an exponential, it never
vanishes
and the value of $\hbar$ could be 
any real number.
We denote the 
line bundle as $(\tilde G_{1+} \times_\rho {\bf C}) /\tilde
B_{1+}$.  A section of this line bundle can be identified with equivariant
functions:
\beq
    \psi:\tilde G_{1+} \to {\bf C} \quad {\rm such \ that \  }
   \psi(\gamma\beta, qr)=\rho(\beta, r)\psi(\gamma, q).
\eeq

Let us exhibit the functions which would satisfy this condition.
They are given by generalized determinants very similar to the case
discussed in \cite{rajtur}, as an example we start with, 
\beq
    \psi(\gamma,q)=\dt_\omega^{1 \over \hbar}(\gamma_{11}q^{-1})
.\eeq
One can see that this is an equivariant function on the 
space $\tilde G_{1+}$, using the properties of the Dixmier trace.
There is no restriction on the value of $\hbar$
due to holomorphicity, since the exponential
is an entire function except possibly due to a topological obstruction.
We can compose more functions of this type, 
if we allow ``mixing'' of the 
elements of $\gamma_{11}$ with the elements of $\gamma_{21}$, in a 
controlled way.
One can see that if we assume that the mixing is allowed by finite
rank operators the result will not change. 
More than that, it will not change if we use elements of $(\ids)^{(0)}$.
To get different functions we need to mix them by elements of
$\ids/(\ids)^{(0)}$.
\beq 
 \dt_\omega((1-A_iS)\gamma_{11}q^{-1}+A_i\gamma_{21}q^{-1})
\eeq
for $A_i \in \ids(\Hi_+,\Hi_-)$ and the mapping 
$S:\Hi_- \to \Hi_+$ is an isometric isomorphism which is given by 
mapping one set of orthonormal basis elements into the other( it can
simply be taken as sending $e_{-i} \to e_i$).
This form is guessed from the system studied in \cite{rajtur,presseg}.
However we have two problems with this form.
When we use 
$A_iS\gamma_{11}$, this expression is not convergent under Dixmier trace.
The second is that we  want our wave functions to be constant when they are 
restricted to the null orbit, as we will see this is not possible for
the above form of the functions due to the term 
$A_iS\gamma_{11}$.
One can see that dropping this term does not change the 
equivariance condition thanks to the Dixmier trace again.
This means an infinite multiplicative renormalization.
At the same time we see that it is simpler to just multiply with the
function $\Tr_\omega(A_i\gamma_{21}q^{-1})$, since it is invariant
under the action of $\tilde B_{1+}$, hence it descends to a function
on the quotient, $\Gr$.
A similar argument shows that we can do better,
we may add even a nonlinear term, which is still invariant.
Any higher order addition vanishes.
We can write down a general expression,
\beq 
 \dt_\omega^{1 \over \hbar}(\gamma_{11}q^{-1})
\Tr_\omega(A_i\gamma_{21}q^{-1})\Tr_\omega(
 B_j\gamma_{21}q^{-1}
 B_k\gamma_{21}q^{-1})
\eeq
where $B_j,B_k \in B(\Hi_+,\Hi_-)$.
For clarity, we will prove that this form is equivariant and satisfies all the
requirements.

Let us look at the action by an element of 
$\tilde B_{1+}$;
\bea
 &\dt_\omega^{1 \over \hbar}&\!\!\!((\gamma\beta)_{11}(qr)^{-1})
\Tr_\omega \big(
A_i(\gamma\beta)_{21}(qr)^{-1})\Tr_\omega(
B_j(\gamma\beta)_{21}(qr)^{-1}
 B_k(\gamma\beta)_{21}(qr)^{-1}\big)=\nn\cr
&=&\!\!\!\dt_\omega^{1 \over \hbar}(\gamma_{11}\beta_{11}r^{-1}q^{-1})
\Tr_\omega(
A_i\gamma_{21}\beta_{11}r^{-1}q^{-1})\Tr_\omega\big(
B_j\gamma_{21}\beta_{11}r^{-1}q^{-1}
 B_k\gamma_{21}\beta_{11}r^{-1}q^{-1}\big)=\nn\cr
&=&\!\!\!\dt_\omega^{1 \over \hbar}(\gamma_{11}(1+I)q^{-1})\Tr_\omega(
 A_i\gamma_{21}(1+I)q^{-1})\Tr_\omega\big(B_j\gamma_{21}(1+I)q^{-1}
 B_k\gamma_{21}(1+I)q^{-1}\big)
\eea
using $\beta_{11}r^{-1}=1+I$ for $I \in \idf$.
\bea
&\dt_\omega^{1 \over \hbar}&\!\!\!(\gamma_{11}q^{-1}(1+qIq^{-1}))
\Tr_\omega(
A_i\gamma_{21}q^{-1}+A_i\gamma_{21}Iq^{-1})
\Tr_\omega((B_j\gamma_{21}q^{-1}+
B_j\gamma_{21}Iq^{-1})\nn\cr
&\times & (B_k\gamma_{21}q^{-1}+B_k\gamma_{21}Iq^{-1}))\nn\cr
&=&\!\!\!\dt_\omega(\gamma_{11}q^{-1}(1+qIq^{-1}))\Tr_\omega\big(
A_i\gamma_{21}q^{-1}\big)
\Tr_\omega\big(B_j\gamma_{21}q^{-1}B_k\gamma_{21}q^{-1})\big)
\eea
In the last line we have used 
$\idf\ids \in {\cal L}^1$ and the Dixmier trace vanishes on them.
As a result we get,
\beq
{\rm exp}{1 \over \hbar}(\Tr_\omega(\gamma_{11}q^{-1}-1)+
\Tr_\omega(\beta_{11}r^{-1}-1))
\Tr_\omega(A_i\gamma_{21}q^{-1})
\Tr_\omega\Big(B_j\gamma_{21}q^{-1}B_k\gamma_{21}q^{-1}\Big)
\eeq
where we use,
$(1+I)(1+qIq^{-1})=1+I+qJq^{-1}+IqJq^{-1}$ and the last term is zero
inside the Dixmier trace for $\gamma_{11}q^{-1}=1+I$, $I \in \idf$.
This is the  equivariance condition we want.
We can take a product over the trace part only, and this will not
change the result.

We need to check  that when we reduce the 
wave function onto the
``small'' orbit, that is to the subgorup 
$\tilde G_{1+}^{(0)}$, the resulting wave functions are just constant.
This is necessary for consistency with the classical Poisson bracket
calculation as we will see.
Let us note the following, in fact our system is invariant
under a larger symmetry.
This will be discussed in the next section, and we will 
see that the correct phase space is smaller.

One can see that we can compose a  product wave function;
in  general we define, as in in the case of $\Di$,
\bea
     \dt_\omega^{1 \over \hbar}(\gamma_{11}q^{-1})\prod_{i,j,k\ }
\Tr_\omega\big(A_i\gamma_{21}q^{-1})
\Tr_\omega(B_j\gamma_{21}q^{-1}
 B_k\gamma_{21}q^{-1} ).
\eea
These and their various superpositions 
are the most general wave functions we can
construct. The next step is to 
complete these set of wave functions with respect to an inner product
to get the quantum Hilbert space.
\footnote{Notice that in our previous work 
\cite{rajtur} we could have used the trace class operators for the 
mixing of $\gamma_{11}$ and $\gamma_{21}$, not only the finite rank
ones.
Various superpositions of the 
wave functions described in that work will
lead to a similar  general form given here with a
finite rank matrix used for the mixing. 
The finite rank operators are dense in the trace class and
the determinant uses the ordinary trace. We can extend the wave
functions in that case to the one  with mixing elements in   
the 
trace class operators.
This should actually be done by using the inner product in the quantum 
Hilbert space, but we expect some dominance property with respect to the 
parameters. In fact the above claim is known to be true,
 hence there is no loss of generality
in that case.}
Notice that still there could be an integrality condition on
$\hbar$, due to the fact that the first Chern class of the line bundle
corresponding to the representation $\rho$ should be in $H^2(\Gr, {\bf
Z})$. 

Let us note that 
the form of the wave function for the Disc and the Grassmanian are 
quite the same--this is unlike the previous case
studied in \cite{rajtur}.
We can in fact set up a 
one-to-one correspondence between the 
elements of the two Quantum Hilbert spaces;
for any choices of  $A_i \in \ids$
,and $B_j, B_k$,which are  operators in
$B(\Hi_-,\Hi_+)$
we have a wave function on the Disc and on The Grassmanian.
This suggests in some natural sense a boson-fermion correspondence,
if we think of the Disc corresponding to a bosonic and 
the Grassmanian to a fermionic system.
Of course this correspondence  is only at a formal level, since we only
have  a set theoretical relation between 
the two Quantum Hilbert spaces.
One also has to check the inner products, to make 
sure that the linearly independent choices are mapped to each other 
in the same manner.
This seems to be a reasonable expectation, its physical 
meaning is not so obvious to the author; it maybe due to the 
fact that we have thrown away a large part of the 
phase space, which may have a lot of physical information about the 
system.
We will see some further evidence of the 
equivalence of these two systems in this quantization scheme,
when we
look at these problems in a different way in the next section.\footnote{there
is larger stability subgroup as we will see in the next section and
it seems to suggest a simpler solution.}

The above wave functions carry a representation of the
group $\tilde G_{1+}$ which comes from the left action;
\beq
  {\rm r}(\gamma',q')\psi(\gamma,q)=
\psi( \gamma'^{-1}\gamma,q^{-1}q')
\eeq
This group action is well-defined. 
We give a proof for completeness:
let us denote the inverse element acting from the left 
by $(\lambda,r) \in \tilde G_{1+}$.
For simplicity we drop the products  
and  compute the following expression:
\beq
\dth_\omega((\lambda\gamma)_{11}q^{-1}r^{-1})
\Tr_\omega(A_i(\lambda\gamma)_{21}q^{-1}r^{-1})
\Tr_\omega(B_j(\lambda\gamma)_{21}q^{-1}r^{-1}
B_k(\lambda\gamma)_{21}q^{-1}r^{-1})
\eeq
We expand the products;
\bea
&\dth_\omega&\!\!\!((\lambda_{11}\gamma_{11}+
\lambda_{12}\gamma_{21})q^{-1}r^{-1})\Tr_\omega(A_i(\lambda_{21}\gamma_{11}+
\lambda_{22}\gamma_{21})q^{-1}r^{-1})\nn\cr
&\times&\Tr_\omega(B_j(\lambda_{21}\gamma_{11}+
\lambda_{22}\gamma_{21})q^{-1}r^{-1}
B_k(\lambda_{21}\gamma_{11}+\lambda_{22}\gamma_{21})q^{-1}r^{-1}))\nn\cr
&=&\!\!\dth_\omega((1+I)(1+rJr^{-1})+\lambda_{12}\gamma_{21}q^{-1}r^{-1})
\Tr_\omega(A_i\lambda_{21}(1+I)r^{-1}+
A_i\lambda_{22}\gamma_{21}q^{-1}r^{-1})\nn\cr
&\times&\!\!\!\Tr_\omega[B_j\lambda_{21}(1+I)r^{-1}+
B_j\lambda_{22}\gamma_{21}q^{-1}r^{-1}]
[B_k\lambda_{21}(1+I)r^{-1}+
B_k\lambda_{22}\gamma_{21}q^{-1}r^{-1}]\nn\cr
&=&\!\!\!\dth_\omega(1+I+J+\lambda_{12}\gamma_{21}q^{-1}r^{-1})\times
\Tr_\omega(A_i(\lambda_{21}r^{-1}+\lambda_{22}\gamma_{21})q^{-1}r^{-1}\nn\cr
&\times&\!\!\!\Tr_\omega[B_j\lambda_{21}r^{-1}+
B_j\lambda_{22}\gamma_{21}q^{-1}r^{-1}]
[B_k\lambda_{21}r^{-1}+B_k\lambda_{22}\gamma_{21}q^{-1}r^{-1}]
\eea
where we use $\gamma_{11}q^{-1}=1+I$ and $\lambda_{11}r^{-1}=1+J$.
Notice also that we used the vanishing of Dixmier trace 
whenever the resulting multiplication is in the trace class
operators--and this is the case for example for products of the 
kind $\lambda_{21}I$ and for other similar terms.
All the terms in the above expression are actually in
the ideal $\idf$ except $1$, hence we have the result 
well-defined.
 
Clearly the action of a group element on a 
general wave function follows directly from this
result.
It follows immediately from the above formula that when we restrict
to the subspace $\tilde G_{1+}^{(0)}$ for the wave function elements
and the 
group element multiplying from the left, the expression inside the
determinant gives us $1$.
This shows that the representation is trivial.

For completeness we will prove that we have a true representation of
the
group $\tilde G_{1+}$ on these space of wave functions.
This is a technical step, and can be skipped.
We will compare the action of the group elements and show that 
\beq
   \psi(\lambda(\sigma\gamma),r(sq))=\psi((\lambda\sigma)\gamma,(rs)q)
,\eeq
hence when we apply the left multiplication on the left 
this gives us a representation.
we have the group action by $(\lambda,r)$ given by the above 
formula that we have already calculated;
\bea
 &\dth_\omega&\!\!\!\!(\lambda_{11}r^{-1})\dth_\omega(\gamma_{11}q^{-1}
+\lambda_{12}\gamma_{21}q^{-1}r^{-1})
\Tr_\omega A_i(\lambda_{21}r^{-1}+\lambda_{22}\gamma_{21}q^{-1}r^{-1})\nn\cr
&\times&\!\!\!\Tr_\omega\Big(B_j(\lambda_{21}r^{-1}+
\lambda_{22}\gamma_{21}q^{-1}r^{-1})B_k
(\lambda_{21}r^{-1}+\lambda_{22}\gamma_{21}q^{-1}r^{-1})\Big)
\eea
We can act with the element $(\sigma,s)$ on this;
it is better to break the terms into separate parts:
\bea
&\dth_\omega&\!\!\!\!(\lambda_{11}r^{-1}) 
\dth_\omega\Big((\sigma_{11}\gamma_{11}+\sigma_{12}\gamma_{21})q^{-1}s^{-1}
+\lambda_{12}(\sigma_{21}\gamma_{11}+\sigma_{22}\gamma_{21})q^{-1}
(rs)^{-1}\Big)\nn\cr
&\dth_\omega&\!\!\!\!(\lambda_{11}r^{-1})\dth_\omega(\sigma_{11}s^{-1})
\dth_\omega\Big(\gamma_{11}q^{-1}+\sigma_{12}\gamma_{21}q^{-1}s^{-1}
+\lambda_{12}\sigma_{21}(rs)^{-1}+\lambda_{12}\sigma_{22}\gamma_{21}
q^{-1}(rs)^{-1}\Big)
.\eea
We compare this with the action of the group element 
$(\lambda\sigma, rs)$ on the same wave function;
\bea
&\dth_\omega&\!\!\!\!((\lambda\sigma)_{11}(rs)^{-1})\dth_\omega(
\gamma_{11}q^{-1}+(\lambda\sigma)_{12}\gamma_{21}q^{-1}(rs)^{-1})=\nn\cr
&\dth_\omega&\!\!\!\!(\lambda_{11}\sigma_{11}s^{-1}r^{-1}+
\lambda_{12}\sigma_{21}s^{-1}r^{-1})\dth_\omega(\gamma_{11}q^{-1}+\lambda_{11}
\sigma_{12}\gamma_{21}q^{-1}s^{-1}r^{-1}+\lambda_{12}\sigma_{22}\gamma_{21}
q^{-1}(rs)^{-1})\nn\cr
&=&\!\!\!\dth_\omega(\lambda_{11}r^{-1})\dth_\omega(\sigma_{11}s^{-1})
\dth_\omega(\gamma_{11}q^{-1}+\lambda_{12}\sigma_{21}(rs)^{-1}
+\sigma_{12}\gamma_{21}q^{-1}s^{-1}+\lambda_{12}\sigma_{22}\gamma_{21}
q^{-1}(rs)^{-1})
.\eea
These two expressions are the same, hence we have a group action on this 
part of the wave function.

Let us check the next term;
\bea
&\Tr_\omega&\!\!\!\!(A_i(\lambda_{21}r^{-1}+
\lambda_{22}(\sigma_{21}\gamma_{11}+\sigma_{22}\gamma_{21})q^{-1}(rs)^{-1})=
\nn\cr
&\Tr_\omega&\!\!\!\!(A_i(\lambda_{21}r^{-1}+
\lambda_{22}\sigma_{21}(rs)^{-1}+\lambda_{22}\sigma_{22}\gamma_{21}q^{-1}
(rs)^{-1})
.\eea
Let us compare this with the direct application of the product;
\bea
&\Tr_\omega&\!\!\!\!(A_i((\lambda\sigma)_{21}(rs)^{-1}+
(\lambda\sigma)_{22}\gamma_{21}q^{-1}
(rs)^{-1})=\nn\cr
&\Tr_\omega&\!\!\!\!(A_i(\lambda_{21}\sigma_{11}s^{-1}r^{-1}+
+\lambda_{22}\sigma_{21}(rs)^{-1}+\lambda_{21}\sigma_{12}\gamma_{21}q^{-1}
(rs)^{-1}+\lambda_{22}\sigma_{22}\gamma_{21}q^{-1}(rs)^{-1})\nn\cr
&=&\!\!\!\!\Tr_\omega(A_i(\lambda_{21}r^{-1}+
\lambda_{22}\sigma_{21}(rs)^{-1}+\lambda_{22}\sigma_{22}\gamma_{21}q^{-1}
(rs)^{-1})
.\eea
These two expressions are the same.
Let us look at the last type of term;
\bea
  &\ &\Tr_\omega\Big(B_j(\lambda_{21}r^{-1}+\lambda_{22}(\sigma_{21}
 \gamma_{11}+
\sigma_{22}\gamma_{21})q^{-1}(rs)^{-1})
B_k(\lambda_{21}r^{-1}+\lambda_{22}
(\sigma_{21}\gamma_{11}+
\sigma_{22}\gamma_{21})q^{-1}(rs)^{-1})\Big)\nn\cr
 &=&\!\!\!\Tr_\omega\Big(B_j(\lambda_{21}r^{-1}+\lambda_{22}\sigma_{21}(rs)^{-1}
+\lambda_{22}\sigma_{22}\gamma_{21}q^{-1}(rs)^{-1})\nn\cr
&\ &B_k(\lambda_{21}r^{-1}+\lambda_{22}\sigma_{21}(rs)^{-1}
+\lambda_{22}\sigma_{22}\gamma_{21}q^{-1}(rs)^{-1})\Big)
.\eea
If we look at the action of the product:
\bea
&\Tr_\omega&\!\!\!\!\Big(B_j(\lambda\sigma)_{21}(rs)^{-1}+(\lambda\sigma)_{22}
\gamma_{21}q^{-1}(rs)^{-1})B_k(\lambda\sigma)_{21}(rs)^{-1}+(\lambda\sigma)_{22}
\gamma_{21}q^{-1}(rs)^{-1})\Big)\nn\cr
&=&\!\!\!\!\Tr_\omega\Big(B_j(\lambda_{21}r^{-1}+
\lambda_{22}\sigma_{21}(rs)^{-1}+(\lambda_{21}\sigma_{12}+
\lambda_{22}\sigma_{22})\gamma_{21}q^{-1}(rs)^{-1})\nn\cr
&\ &B_k(\lambda_{21}r^{-1}+
\lambda_{22}\sigma_{21}(rs)^{-1}+(\lambda_{21}\sigma_{12}+
\lambda_{22}\sigma_{22})\gamma_{21}q^{-1}(rs)^{-1})\Big)\nn\cr
&=&\!\!\!\!\Tr_\omega\Big(B_j(\lambda_{21}r^{-1}+
\lambda_{22}\sigma_{21}(rs)^{-1}
+\lambda_{22}\sigma_{22}\gamma_{21}q^{-1}(rs)^{-1})\nn\cr
&\ &B_k(\lambda_{21}r^{-1}+\lambda_{22}\sigma_{21}(rs)^{-1}
+\lambda_{22}\sigma_{22}\gamma_{21}q^{-1}(rs)^{-1})\Big)
,\eea
where we used the vanishing of Dixmier trace on ${\cal L}^1$ 
and some rearrangements on all the above calculations.
We see that this term also respects the group action.

The above representation  factors through the subgroup 
which corresponds to 
the elements $s$ of $GL(\Hi_-)$, with $\dt_\omega(s)$ well-defined.
These are the elements in the subgroup,
$s \in 1+\idf$.
Notice that for a fixed element $\gamma$, the freedom we have to
choose
different $q$'s which satisfy the determinant condition, is 
isomorphic to $GL^{1+}=\{q \in GL(\Hi_-)| q=1+\idf\}$
Although we could not find a way  to reduce it to this group everywhere.
Hence our representations can actually be reduced to the
representations of another group, which is a central extension of the 
group $GL_{1+}$. The extension is trivial on the subgroup 
$GL_{1+}^{(0)}$, this can be seen by noticing 
that $s \in 1+(\idf)^{(0)}$ will give us $\dt_\omega(s)=1$.
We can construct a commutative diagram as we have done in 
\cite{rajtur}, to show that this gives us a
representation of a central extension of the 
group $ GL_{1+}$.
This is more transparent if we look in the Lie algebra level, and also
this gives us a chance to compare the central term we have in the case
of moment maps.
The Lie algebra we need to consider is clearly 
$\{(u,r)|\ \ r \in {\rm End}(\Hi_-)\quad u_{11}-r \in \idf\}$.
One can see that the Lie bracket
$[(u,r),(v,s)]=([u,v],[r,s])$ is well-defined, and this is the
infinitesimal form of the group $\tilde G_{1+}$.
One can see immediately that the set of possible $r$'s is isomorphic
to
the set of $r \in GL(\Hi_-)$ such that $r \in \idf$.
We can construct a central extension of the original  Lie algebra by using,
$(u,r) \mapsto (u, \Tr_\omega(u_{11}-r))$.
We give the trivial Lie bracket to the complex numbers.
Under this map,  the commutator goes to
\beq
([u,v],\Tr_\omega([u,v]_{11}-[r,s]))=
\big([u,v],\Tr_\omega^\epsilon([\epsilon,u]v)\big)
\eeq
where we used
\bea
&\ &\Tr_\omega(u_{11}v_{11}-v_{11}u_{11}+u_{12}v_{21}-
v_{12}u_{21}-[r,s])\nn\cr
&\ &=\Tr_\omega(u_{11}v_{11}-rs-v_{11}u_{11}+sr)+\Tr_\omega(u_{12}v_{21}-
v_{12}u_{21})\nn\cr
&\ &=\Tr_\omega((u_{11}-r)v_{11}+r(v_{11}-s))-
\Tr_\omega(v_{11}(u_{11}-r)+(v_{11}-s)r)+\Tr_\omega^\epsilon([\epsilon,u]v)
\nn\cr
&\ &=\Tr_\omega^\epsilon([\epsilon,u]v)
\eea
since all the terms, as grouped, are in $\idf$ and we use the 
Dixmier trace properties.
Note that this is the same central extension in the case of 
moment maps that we have been using.
We will show by an explicit calculation that 
our representations can be reduced to representations of this algebra;
let us first compute the infinitesimal action, for simplicity
we drop the product signs: 
\beq
 {\cal L}_{(u,r)}\dth_\omega(\gamma_{11}q^{-1})\Tr_\omega(
A_i\gamma_{21}q^{-1})\Tr_\omega(B_j\gamma_{21}q^{-1}
 B_k\gamma_{21}q^{-1})
\eeq
It is again easier to use the derivation property of the 
Lie derivative and compute individual 
terms.
\bea
&{\cal L}_{(u,r)}&\!\!\!\!\dth_\omega(\gamma_{11}q^{-1})=
{-1 \over\hbar}\Tr_\omega\Big((u_{11}\gamma_{11}+u_{12}\gamma_{21})q^{-1}
-\gamma_{11}q^{-1}r)\dth_\omega(\gamma_{11}q^{-1})\nn\cr
&=&\Big(-{1 \over \hbar}\Tr_\omega(u_{11}-r)-{1 \over \hbar}
\Tr_\omega(u_{12}\gamma_{21}q^{-1})\Big)\dth_\omega(\gamma_{11}q^{-1})
.\eea
In a similar way we get;
\bea
&{\cal L}_{(u,r)}&\!\!\!\!\Tr_\omega(A_i\gamma_{21}q^{-1})=
\Tr_\omega(A_i(u_{21}+u_{22}
\gamma_{21}q^{-1}-\gamma_{21}q^{-1}r))
\eea
and
\bea
&{\cal L}_{(u,r)}&\!\!\!\!\Tr_\omega\Big(B_j\gamma_{21}q^{-1}
B_k\gamma_{21}q^{-1}\Big)\nn\cr
&=&\Tr_\omega\Big(B_j(u_{21}+u_{22}\gamma_{21}q^{-1}-\gamma_{21}q^{-1}r)B_k
(u_{21}+u_{22}\gamma_{21}q^{-1}-\gamma_{21}q^{-1}r)\Big)
.\eea
We  define  a new representation, and using the above expressions check that 
it is in fact independent of the choice of the Lie algebra  element
$r$;
\beq
 \hat {\rm r}[(u,r)]={\cal L}_{(u,r)}+{1 \over \hbar}\Tr_\omega(u_{11}-r)
,\eeq
acting on the same set of wave functions.
Using the above expression, we compute the 
action of the Lie algebra element 
$(u,r+s)$, where $s \in \idf$ and this is the freedom we have.
It is again simpler to check this on each basic piece;
\bea
(\hat {\rm r}[(u,r+s)]&\dth_\omega&\!\!\!\!(\gamma_{11}q^{-1}))\phi(A_i,B_j,B_k,\gamma,q)\nn\cr
&=&\!\!\!\![{\cal L}_{(u,r+s)}+
{1 \over \hbar}\Tr_\omega(u_{11}-(r+s))]
\dth_\omega(\gamma_{11}q^{-1})\phi(A_i,B_j,B_k,\gamma,q)\nn\cr
&=&\!\!\!-{1 \over \hbar}\Tr_\omega(u_{12}\gamma_{21}q^{-1})
\dth_\omega(\gamma_{11}q^{-1})\phi(A_i,B_j,B_k,\gamma,q)
.\eea
For the other terms we only use the Lie derivative part since the 
scalar part is used in the above expression already;
\bea
&\ &{\cal L}_{(u,r+s)}\Tr_\omega(A_i\gamma_{21}q^{-1})= 
 \Tr_\omega(A_i(u_{21}+u_{22}
\gamma_{21}q^{-1}-\gamma_{21}q^{-1}(r+s)))\nn\cr
&=&\Tr_\omega(A_i(u_{21}+u_{22}\gamma_{21}q^{-1}-A_i
\gamma_{21}q^{-1}r)={\cal L}_{(u,r)}\Tr_\omega(A_i\gamma_{21}q^{-1})
,\eea
by using the fact that everytime $s$ is multiplied with an element,
the resulting term is in the ideal of trace class operators, and the
Dixmier  trace vanishes on them.
The other term,
\beq
{\cal L}_{(u,r+s)}\Tr_\omega(B_j\gamma_{21}q^{-1}B_k\gamma_{21}q^{-1})
,\eeq
can be shown to be independent of $s$ by using the same reasoning as above.
Hence we can denote the representation we have as 
$\hat{\rm r}(u)$.

If we compute the commutator,
\bea
 (\hat {\rm r}(u)\hat{\rm r}(v)-\hat{\rm r}(v)\hat {\rm r}(v))\psi(\gamma,q)=
[\hat {\rm r}([v,u])-{1 \over \hbar}\Tr_\omega([u,v]_{11}-[r,p])]\psi(\gamma,q)
.\eea
The last term is independent of the choices of $r,p$ and 
equal to $-{1 \over \hbar}\Tr_\omega^\epsilon[\epsilon,u]v$ 
as we have seen before.

Hence the representations that we have obtained can be reduced to the
representations of the central extension of the 
Lie group $U_{1+}(\Hi)$.
This is the quantization of our classical system,
it may not be possible  to express the central term corresponding 
to the group in this form, since the extension may have  a topological
twist in general.
We are not able to answer this question, although the discussion 
in the last section hints that the correct phase space is topologically 
trivial. This will imply that the central extension actually comes from 
a central term, globably defined.

\section{Flat Geometry and Quantization}

In this section we will introduce  a classical system 
which appears to be unrelated  at first sight.
This point of view was suggested  by Rajeev in our discussions.
We consider the set of elements $\Z$ such that 
they belong to the following quotient space 
$\idsf /(\idsf)^{(0)}$
 (equivalence classes of $Z :\Hi_+ \to \Hi_-$ and 
$Z \in \idsf$ under $Z-Z' \in (\idsf)^{(0)}$).
There is a natural quotient norm;
\beq
   |||\Z|||=\inf_{Z_0 \in (\ids)^{(0)}}||Z+Z_0||_{\ids}
\eeq
where $Z$ is a representative in the equivalence class of $\Z$.

There is a natural product from 
$\ids \times  \ids \to \idf$, and this reduces to the 
quotients;
$\ids /(\ids)^{(0)} \times \ids /(\ids)^{(0)} \to 
\idf / (\idf)^{(0)}$ given by 
$\Z\ \Z' =\overline{ZZ'}$
The natural product $B(\Hi) \times \ids \to \ids$ 
also descends to the quotient;
$B(\Hi) \times (\ids/(\ids)^{(0)}) \to \ids/(\ids)^{(0)}$.

The Dixmier trace is  
nondegenerate on this quotient space. 
We will use this important fact to introduce 
an obvious symplectic form.
\beq
    \tilde \Omega_\omega=i\Tr_\omega d\Z \wedge d \Z^\dag
\eeq

This flat geometry has a simple symmetry group\footnote{This may not be the 
most general action, but it is the obvious one.}; given by 
rotations   and  translations. Due to the quotient  we can allow for 
slight deviations from unitary transformations
and  write a general 
transformation as
\beq
  \Z \mapsto \overline{\overline{eZf^{-1}}+l}
,\eeq
where $e \in GL(\Hi_-),\ f \in GL(\Hi_+)$ such that 
$e^\dag e-1, f^\dag f-1 \in K(\Hi)$, and 
$ l \in \ids $. One can check that this is in fact a group under the 
obvious composition law, which we call as the affine group,
${\cal A}_{1+}$.
This action is well-defined and transitive. 
One can immediately check that the group action 
preserves the symplectic form due to the extra conditions we have;
\bea
 \tilde \Omega_\omega=i\Tr_\omega e d \Z f^{-1} \wedge 
f^{-1\dag} d\Z^\dag e^\dag=i\Tr_\omega(e^\dag e)d\Z \wedge 
(f^\dag f)^{-1} d\Z^\dag=\tilde \Omega_\omega 
,\eea
using the fact that the porduct $e^\dag e d\Z \approx d\Z$
and the same for $f$(see Appendix for a proof).

It is again natural to find the moment maps generating this  
action. We can find them using the infinitesimal form of the 
group action;
$\tilde V_{(e,f,l)}(\Z)=\overline{ \alpha Z}-
\overline{Z \delta}+\overline\beta$(here, we denote the 
Lie algebra elements by the same letter $\alpha, \beta, \delta$, but
they now satisfy $\alpha^\dag+\alpha=1+K, \delta^\dag+\delta=1+K$ where
$K$ is a compact operator, and $\overline\beta \in \ids/(\ids)^{(0)}$,
yet we denote the moment maps by $F_{(e,f,l)}$ to imply that they are
coming from the affine action. we hope that this does not cause too much 
confusion).
Hence,
\beq
   F_{(e,f,l)}=i\Tr_\omega(\alpha \Z \ \Z^\dag-
    \Z \delta \Z^\dag+\overline\beta \ \Z^\dag+
\Z \ \overline \beta^\dag)
.\eeq
If we compute 
$\{ F_{(e_1,f_1,l_1)}, F_{(e_2,f_2,l_2)}\}$ we will see that there is 
a central term. Since we will do this calculation below to make
connection with the previous section, we postpone the result.

Geometric quantization gives us immediately the following 
general set of  wave functions;
\beq
 \Psi(\Z, \Z^\dag)=e^{-{1 \over \hbar}\Tr_\omega \Z \ \Z^\dag}   
 \prod_{i, j,k } \Tr_\omega(\bar A_i \Z)\Tr_\omega
(B_j \Z B_k \Z)
,\eeq
where $\bar A_i \in \ids$ and $B_{j,k}$ are bounded.
Naturally, this set of wave functions carry a 
representation of the central extension of the above group action,
via the same type of operators we have found before\footnote{We 
skip  a detailed derivation of this formula, but the reader can verify it by 
using standard geometric quantization.};
\beq
  \hat F \psi(\Z)=({\cal L}_{(\alpha, \delta,\beta)}
-{1 \over \hbar}\Tr_\omega(\overline \beta^\dag \Z))\psi(\Z)
.\eeq

This system has an interesting 
connection to our discussions on the previous section.
Let us recall the Disc
case. One can recover the Symplectic form for the Disc using the 
following K\"ahler potential, 
$i\Tr_\omega \log (1-ZZ^\dag)$, just as in the finite dimensional
case.
Let us expand the 
K\"ahler form, and use the properties of the 
Dixmier trace. We see that the result is a simple expresion;
$i\Tr_\omega Z Z^\dag$.
This is the result for a flat system, except for degeneracies.
If we look at the quotient, as above, the result is the same as the 
K\"ahler potential of the above system.

We can also apply the quotient homomorphism to 
 our pseudo-unitary group action; this gives,
\beq 
 \bar Z \mapsto \overline{aZd^{-1}}+\overline{bd^{-1}}
.\eeq
Let us show that the group property is preserved under this mapping;
\bea
  \overline{g_2 \circ (g_1 \circ Z)}&=&
  a_2(a_1\Z d_1^{-1}+\overline{b_1d_1^{-1}})d_2^{-1}+
\overline{b_2d_2^{-1}}=
  a_2a_1\Z (d_2d_1)^{-1}+a_2\overline{b_1(d_2d_1)^{-1}}=\nn\cr
  &=&\overline{(a_2a_1+b_2b_1)Z (d_2d_1+c_2c_1)^{-1}}+\overline{(a_2b_1+b_2d_1)
 (d_2d_1+c_2c_1)^{-1}}=\overline{(g_2g_1) \circ Z}
\eea
This gives us an embedding of $U_{1+}(\Hi_-,\Hi_+)$ into the affine group.
Another interesting point is to look at the moment maps; and expand
$\Tr_\omega u(\Phi-\epsilon)$ in the variable $Z$, by using the 
expression of $\Phi$ in terms of $Z$.
The properties of the Dixmier trace can be used to see that most 
of the terms vanish;
the result is the same as the moment maps of the flat system:
\beq
  f_u=i\Tr_\omega(\alpha ZZ^\dag-Z\delta Z^\dag+\beta Z+Z^\dag
\beta^\dag)
\eeq
Of course, it is natural to go to the quotient again, and we get
$F_u=F_{(a,d,bd^{-1})}$.
We can now compute to see the Poisson bracket of these two moment
maps, using the flat Poisson bracket;
\bea
  \{ F_u, F_v\}&=&i\Tr_\omega([\alpha_1,\alpha_2]\Z\ \Z^\dag-
 \Z[\delta_1,\delta_2]\Z^\dag+
 (\alpha_2\beta_1-\alpha_1\beta_2+\beta_1\delta_2-\beta_2\delta_1)\Z^\dag\nn\cr
 &+&\Z(\alpha_2\beta_1-\alpha_1\beta_2+\beta_1\delta_2-\beta_2\delta_1)^\dag)
 +i\Tr_\omega(\overline{\beta_1\beta_2^\dag-\beta_2\beta_1^\dag})
.\eea
One can verify that the last term is a central term which is 
equal to the central term we have found before.

The above set of wave functions are equivalent to the wave functions
on the disc and they carry the same representation of the 
central extension of the quotient group.
This shows that the system we studied without the reduction 
can be put into a slightly bigger  flat system.
 
The same question then arises for the Grassmannian. Its
coordinatization will show that in each coordinate neighborhood, the
symplectic form is given by the flat one, and 
similarly the moment functions will look like the flat geometry.
Certainly, the quotient point of view, using 
$U_{1+}/(U_{1+})^{(0)}$, implies that there is a similar
simplification.
Now we will try to present an alternative point of view in the Grassmanian 
which keeps the complex structure.
Consider the following subgroup;
\beq 
  \tilde G_{(1+,0)}=\Big\{(g,q)| g \in 
\pmatrix{{\cal B}&\ids\cr ({\ids})^{(0)}&{\cal B}}\Big\}
. \eeq
This is a closed subgroup, hence the quotient is a holomorphic manifold.
Notice that the representation we have introduced for the 
subgroup ${\tilde B}_{1+}$, actually extend to a representation of this 
larger group:
\beq
   \dth_{\omega}((\gamma\lambda)_{11}s^{-1}q^{-1})=
  \dth_\omega(\gamma_{11}\lambda_{11}s^{-1}q^{-1}+\gamma_{12}\lambda_{21}s^{-1}
q^{-1})
,\eeq
notice that the last term is actually zero under the Dixmier trace 
and the rest follows as before, showing that it is a one-dimensional 
holomorphic representation.
The next thing is to check that the 
remaining part of the wave function is in fact invariant under 
the group $\tilde G_{(1+,0)}$. Thus, we will have a line bundle on this 
smaller quotient, which is the physically relevant phase space.
Let us only check one of the terms;
\bea
   &\ &\Tr_\omega(A_i(\gamma\lambda)_{21}s^{-1}q^{-1})=
   \Tr_\omega(A_i(\gamma_{21}\lambda_{11}s^{-1}q^{-1}+\gamma_{22}\lambda_{21}
s^{-1}q^{-1}))\nn\cr
  &\ &=\Tr_\omega(A_i\gamma_{21}q^{-1})
,\eea
by using the fact that the last term is in $(\idf)^{(0)}$.
Similarly for the other type of term.
Hence we can consider the set of functions as the setions of a 
line bundle on the quotient;
$\tilde G_{1+} / \tilde G_{(1+,0)}$.
Since we already know that 
for $B_{1+}$ the quotient cancel out the $q$ parts, for the above form also we 
get 
\beq
\tilde G_{1+} / \tilde G_{(1+,0)}\approx GL_{1+}/GL_{(1+,0)}=
\pmatrix{{\cal B}&\ids\cr \ids &{\cal B}}/
\pmatrix{{\cal B}&\ids\cr (\ids)^{(0)} &{\cal B}}\approx \ids/(\ids)^{(0)}.
\eeq
This shows that the relevant part of the phase spaces for the 
Disc and the Grassmanian are of equal size.
We are not able to provide a link with this and the 
coordinate description at the present moment.
Our guess is that Grassmanian also has the same embedding into a 
flat system, this manifests the possible equivalence of the two systems. 
We hope to clarify some of these issues in a future publication.

\section{Acknowledgements}
Most of this paper owes a lot to several conversations with 
S.G. Rajeev, and shows his guidance and influence. We thank J. Mickelsson
for reading and suggesting corrections, 
and to M. Wodzicki for a series of marvelous lectures 
on traces, to E. Langmann for several suggestions.\footnote{Unfortunately,
we could not incorporate his suggestions into this version.}
We also would like to thank J. Brodzki, P. Bongaart, 
K. Gawedzki, G. Landi,  
M. Walze, 
for many  useful discussions. 
The author gratefully acknowledges the hospitality of
S. Majid and his group, and the kind invitation of P. Goddard and
D. Crighton to DAMTP, where most of this work done during the
author's stay in Cambridge.
During this project the author is supported by the 
EPDI grant, and we thank  IHES for the excellent 
help and working environment provided.

\section{Appendix}

The definitions of the operator ideals will be given.
Let us start with  the definition of $\ids$.
Operator ideals contains compact operators, thus they are 
given by the summability properties of the singular values of the 
operators.
If $s_n(A)=n$th eigenvalue of $|A|$,
then we define a new norm:
\beq
     ||A||_{\ids}=\sup_{N}{\sum_{n=1}^{N} s_n(A) \over \sum^{N}_{i=1}
     {1\over n^{1/2}}}
\eeq
The set of all $A\in K(\Hi)$ for which the above norm is 
finite is  denoted by $\ids$.
It is a symmetrically normed ideal \cite{gohberg}.
Since the sequence ${1 \over n^{1/2}}$ is regular, 
the same ideal can be defined through the asymptotic behaviour of the 
singular values.
In fact, the set of operators in $\ids$ can be defined as 
$A \in K(\Hi)$ such that 
$s_n(A)=O(n^{-1/2})$.
This also gives a simple characterization of the 
completion of the finite rank operators inside $\ids$, denoted as 
$(\ids)^{(0)}$;
$A \in (\ids)^{(0)}$ iff $s_n(A)=o(n^{-1/2}).$\footnote{The symbol
$s_n=O(\pi_n)$ means that, $\limsup_{n \to \infty}{s_n \over \pi_n}<\infty$ and
 $s_n=o(\pi_n)$ iff $\lim_{n \to \infty}{s_n \over \pi_n}=0$.}
One can define the norm for $\idf$ in the same way replacing the 
sequence $1/n^{1/2}$ by $1/n$.
This is not a regular sequence so the completion of finite rank operators 
are given by the behaviour of the partial sums;
$\sigma_N(A)=\sum_{n=1}^{N}s_n(A)$, $\sigma_N(A)=O(\log N)$.
If $s_n(B)=o({1 \over n})$, then it implies that 
$\sigma_N(B)=o(\log N)$, hence $B \in (\idf)^{(0)}$.
But the converse is not true.

We give a  proof that $A, \ B \in \idf$ then $AB \in {\cal L}^1$.
Let us assume that the hypothesis  is true.
It implies that $A,\ B \in {\cal L}^2$ as well.
But we know that ${\cal L}^2 {\cal L}^2 \in {\cal L}^1$,
hence the result.

Next, we will prove that 
if $A \in \ids$ and $B \in (\ids)^{(0)}$ then
$AB \in (\idf)^{(0)}$.

We will  use the inequalities satisfied by
 the singular values. 
\beq
    \mu_{n+m}(AB) \leq \mu_n(A) \mu_m(B)
\eeq
Choose $n+m=2N+j$ where $j=0,1$ and look at the following limit, 
\beq 
    \limsup_{N \to \infty} (2N+j)\mu_{2N+j}(AB)\leq \limsup_{N \to
 \infty }
(2N+j)^{1/2}\mu_N(A)(2N+j)^{1/2}\mu_{N+j}(B). \label{inq1}
\eeq
Now we can use, 
\beq 
   \lim_{N \to \infty} (N^{1/2}\mu_N(A))=a <\infty
 \ \ {\rm and} \ \ \lim_{N \to \infty} ((N^{1/2}\mu_N(B))=0
\eeq
in the above expression to get,
\beq
 \lim_{N \to \infty} (2N+j)\mu_{2N+j}(AB)=0
\eeq
and this implies that the product is in the closure, 
$AB \in (\idf)^{(0)}$.
One can imitate the above proof  to show  that  
$A,\ B \in \ids$ then 
$AB \in \idf$; we leave this to the reader.

Let us use the same idea to show that
$K(\Hi)\ids \subset (\ids)^{(0)}$;
\beq
 \limsup_{2N+j}(2N+j)^{1/2}\mu_{2N+j}(AK) \leq \limsup_{2N+j}
  (2N+j)^{1/2}\mu_{N}(A)\mu_{N+j}(K)=0
 ,\eeq
by using the fact that $\lim_{N} \mu_N(K)=0$ for a compact operator $K$.

In the second part we will prove some of the 
properties of the conditional Dixmier traces, which are identical to the 
usual trace  conditions.
We define the conditional trace as
\beq
     \Tr_\omega^\epsilon A=\Tr^\epsilon_\omega \pmatrix{ a_{11} & a_{12}\cr 
    a_{21} &a_{22}}={1 \over 2}(\Tr_\omega(a_{11})+\Tr_\omega(a_{22}))={1 \over
4}\Tr_\omega[A+\epsilon A \epsilon].
\eeq
for $A=\pmatrix{ a_{11} & a_{12}\cr 
    a_{21} &a_{22}}$. Notice that we have absorbed a factor of $1/2$
into 
the definition
to make the formulae involving this trace look simpler.
First property;
\beq
    \Tr^\epsilon_\omega AB=\Tr^\epsilon_\omega BA
\eeq
if all the individual terms in the products $\sum_k a_{ik}b_{ki}$ are in the
ideal $\idf$.
This is easy to see if we use 
the definition of the conditional trace, and,
\beq
\pmatrix{ a_{11} & a_{12}\cr 
    a_{21} &a_{22}} \pmatrix{ b_{11} & b_{12} \cr b_{21} & b_{22}}=
   \pmatrix{ a_{11}b_{11}+a_{12}b_{21} & \ *\ \cr \ * \ &
a_{21}b_{12}+a_{22}b_{22}}
\eeq
If each of the terms appearing in the diagonal parts are indepently in 
$\idf$, we can use $\Tr_\omega a b=\Tr_\omega b a$ and see that 
the result is the same when one takes the product in the opposite
order.

We will show that the group representation is satisfied upto a central
term.
It is more convenient to use  
${\rm det}_\omega$ instead of $e^{\Tr_\omega}$ and the power $1 / \hbar$
is not written since it is easy to put back.
We compare the two ways of applying the representation, 
$\rho(g_2^{-1})\rho(g_1^{-1})\Psi(Z)$ and
$\rho((g_1g_2)^{-1})\Psi(Z)$.
The first expression gives,
\beq
  {\rm det}_\omega(d_1^{-1}c_1(a_2 Z+b_2)(c_2 Z + d_2)^{-1}+1)
{\rm det}_\omega(d_2^{-1}c_2 Z +1)\Psi((g_1g_2) \circ Z)
\eeq
and the second 
\beq
  {\rm det}_\omega((d_1d_2+c_1b_2)^{-1}(c_1a_2+d_1c_2)Z
+1)\Psi((g_1g_2)^{-1} \circ Z)
.\eeq
It is enough to compare the ``determinant'' pieces
because the other parts are the same.
Let us check the following,
\bea
\ \ \ &{\rm det}_\omega &\!\!\!(d_1^{-1}c_1(a_2 Z+b_2)(c_2 Z + d_2)^{-1}+1)
\dt_\omega(d_2^{-1}c_2 Z
+1)\cr \nn
&=&\dt_\omega(d_2^{-1}d_1^{-1}c_1(a_2Z+b_2)(d_2^{-1}c_2Z+1)^{-1}
+1)\dt_\omega(d_2^{-1}c_2 Z
+1)\cr\nn
&=&
\dt_\omega((d_1d_2)^{-1}c_1(a_2Z+b_2)(d_2^{-1}c_2Z+1)^{-1}(d_2^{-1}c_2Z+1)+
d_2^{-1}c_2Z+1)\cr \nn
&=&\dt_\omega((d_1d_2)^{-1}c_1(a_2Z+b_2)+d_2^{-1}c_2Z+1)\cr \nn
&=& \dt_\omega((d_1d_2)^{-1}(c_1a_2+d_1c_2)Z+(d_1d_2)^{-1}c_1b_2+1)
.\eea
We used the multiplicative property of the $ \dt_\omega$, which comes
from the properties of the Dixmier trace. The equalities are true by
adding terms which give zero under dixmier trace, this is the
advantage of using the symbol $\dt_\omega$.
let us compare this with
\bea
 \ \ &{\rm det}_\omega &\!\!((d_1d_2+c_1b_2)^{-1}(c_1a_2+d_1c_2)Z
+1)=\cr \nn
&\quad =&\dt_\omega[((d_1d_2)^{-1}c_1b_2+1)^{-1}
(d_1d_2)^{-1}(c_1a_2+d_1 c_2)Z+1)\cr\nn
&\quad =&\dt_\omega((d_1d_2)^{-1}c_1b_2+1)^{-1}
\dt_\omega((d_1d_2)^{-1}(c_1a_2+d_1c_2)Z+(d_1d_2)^{-1}c_1b_2+1)
\eea
hence they differ by a constant multiple, which never vanihes, 
\beq
 c_\omega(g_1,g_2)=\dt_\omega((d_1d_2)^{-1}c_1b_2+1)
.\eeq
This trace is well-defined as one can check and since it never
vanishes the two sided are equal.
We need to further check that it obeys the cocyle condition,
\bea
    c_\omega(g_1g_2,g_3)c_\omega(g_1,g_2)&=&c_\omega(g_1,g_2g_3)
  c_\omega(g_2,g_3)\nn\\
     &=&{\rm det}^{1 \over \hbar}[(d_1d_2d_3)^{-1}c_1a_2b_3
       +(d_2d_3)^{-1} c_2b_3+(d_1d_2)^{-1}c_1b_2+1].
\eea
The sum of all the terms except $1$ inside the determinant sign are in the ideal
$\idf$, hence the Dixmier trace is well defined.
The cocycle $c_\omega$, in the finite dimensional case, can be
obtained from $\phi(g)=$det$(d)$, as 
$c(g_1,g_2)=\phi(g_1)\phi(g_2)\phi(g_1g_2)^{-1}$. Clearly
it  is not well-defined in  infinite dimensions; in fact,
the extension is nontrivial. Thus, we obtain a
representation of
a central extension $\hat{U}_{1+}({\cal H}_-,{\cal H}_+)$ in
the Quantum Hilbert space of holomorphic sections.

\section{small appendix}
This appendix does not belong to the actual paper, since we could not find a 
reference for the direct proof of the claim in the footnote, 
we include it for those who are
interested in. It is certainly known and 
implicit( or stated in a more general context)
 in the paper of Connes ``non-commutative 
geometry'' IHES Publ. Math. vol. 62, pg 257.

Here we will give the proof that the general linear group is a 
topological product of unitary times positive hermitian elements:

We remind the reader the following formula for the square root of 
a positive element:
\bea
    A^{1/2}={1 \over \pi}\int_{0}^{\infty}{d\lambda \over \lambda^{1/2}}
   A(\lambda I+A)^{-1}
\eea
We need to show that under the new norm
$||[\epsilon, A]_+||+||[\epsilon, A]||_{\ids}=
||a||+||d||+||b||_{\ids}+||c||_{\ids}$, 
the operation is well defined and continuous.
First let us show that it is well-defined.
we can compute first;
\bea
  ||[\epsilon, A^{1/2}]||_{\ids}\leq
 {1 \over \pi}\int_{0}^{\infty}d\lambda\lambda^{1/2}
   ||[\epsilon, (\lambda I+A)^{-1}]||_{\ids}
\eea
using $A(\lambda I+A)^{-1}=1-\lambda(\lambda I+A)^{-1}$.
We can use now,
$[\epsilon,(\lambda I+ A)^{-1}]=-(\lambda I+A)^{-1}[\epsilon, A]
(\lambda I+A)^{-1}$, and the symmetric norm property;
\bea
  ||[\epsilon, A^{1/2}]||_{\ids}&\leq&
 {1 \over \pi}\int_{0}^{\infty}d\lambda\lambda^{1/2}
  ||(\lambda I+A)^{-1}||^2
   ||[\epsilon, A]||_{\ids}\cr\nn
  &\leq&  {1 \over \pi}\int_{0}^{\infty}d\lambda\lambda^{1/2}
  {1 \over (\lambda+{\rm inf}\sigma(A))^2}
   ||[\epsilon, A]||_{\ids}
\eea
Here we used the fact that the spectrum satisfies $\sigma((\lambda I+A)^{-1})=
(\sigma(\lambda I+A))^{-1}$. This shows that the spectrum is contained in
$[(\lambda +||A||)^{-1},(\lambda+{\rm inf}\sigma(A))^{-1}]$.
The infimum of the spectrum of $A$, is a positive number;
hence the integral on the right is actually convergent in the 
$\lambda \approx 0$ region, and as for the $\lambda \to \infty$,
it is clear that the integral is dominated by the integral of 
$\lambda^{-3/2}$, hence convergent. So we have a finite number 
times the norm of the off-diagonal parts, which are finite.
For the diagonal parts we know that the square root is a 
bounded operator\cite{bratelli}, and we use
$||P_+BP_+|| \leq ||B||$, same for $||P_-BP_-||$, 
 for any bounded operator $B$, where $P_+,P_-$ are projections onto
$\Hi_+,\Hi_-$ respectively.
This shows that each one is bounded.

For the continuity, we use similar ideas;
first prove this for bounded operators:
\bea
  ||A^{1/2}-B^{1/2}|| &\leq&  {1 \over \pi}\int_{0}^{\infty}
     d\lambda\lambda^{1/2}
   || (\lambda I+A)^{-1}-(\lambda I+B)^{-1}||\cr\nn
   &\leq& {2 \over \pi}\int_{0}^{\infty}d\lambda\lambda^{1/2}
||(\lambda I+A)^{-1}||^2 ||A-B||\leq C||A-B||.
\eea
Now we can apply a similar method for the off-daigonal parts;
\bea
  ||[\epsilon, A^{1/2}-B^{1/2}]||_{\ids}&\leq&\cr\nn
 {1 \over \pi}\int_{0}^{\infty}&d\lambda&\!\!\!\lambda^{1/2}
  2||(\lambda I+A)^{-1}[\epsilon,A](\lambda I+A)^{-1}-
   (\lambda I+B)^{-1}[\epsilon, B](\lambda I+B)^{-1}||_{\ids}\cr\nn
  &\leq&  {1 \over \pi}\int_{0}^{\infty}d\lambda\lambda^{1/2}
  \Big(||(\lambda I+B)^{-1}||^2||[\epsilon,A-B]||_{\ids}\cr\nn
 &+&\!\!2||(\lambda I+B)^{-1}||^3||[\epsilon,A]||_{\ids}||A-B||\Big)
,\eea
in a small neighborhood of  
$A-B$--that is 
for 
$||[\epsilon,A-B]_+||+||[\epsilon,A-B]||_{\ids}$ sufficiently
small. All  the integrals are convergent and the continuity is clear from this.
SInce $A \mapsto A^\dag A$ is continous under the same topology,
the polar decomposition is a continous operation in this ideal.
We remark that the above proof remains valid for any symmetrically normed 
ideal.

Let us also comment on the proof that $GL_{1+}$ is contractible 
to its unitary subgroup.
Define the polar decomposition, 
$A=U_A|A|$, where $|A|=(A^\dag A)^{1/2}$. 
We have shown that the absolute value map is continuous in the 
topology we use.
Then we can obtain a homotopy to the unitary part, by using
\beq
   A(t)=U_A \exp(t\log|A|)
.\eeq
This gives $A(1)=A$ and $A(0)=U_A$.
Of course we need to justify that the above functions are all continuous
in the product topology.
This can be done by using the following integral representation for 
$\log A$ for positive $A$;
\beq 
 \log A=(A-I)\int_{0}^{1} ds((1-s)I+sA)^{-1}
.\eeq

One can give a self-contained proof by using a very similar 
idea as in the first case.
For this we need to employ a similar identity for the 
fractional powers of positive operators:
\beq
     A^\alpha={\sin \alpha \pi \over \pi}\int_{0}^{\infty}d\lambda
\lambda^{\alpha-1} (\lambda 1+ A)^{-1}A 
,\eeq
 for $0 < \alpha <1$.
By using exactly the same types of ideas we can prove that 
the map $GL_{1+}(\Hi) \times [0,1] \to U_{1+}(\Hi)$ is continuous, hence
$U_{1+}$ is   a deformation retract of $GL_{1+}$.
We give a sketch here.
Let us estimate the norm of $A^\alpha$;
\beq
    ||A^\alpha|| \leq {|\sin \alpha \pi |\over \pi}
    \int_{0}^{\infty}d\lambda \lambda^{\alpha -1}||A||(\lambda+||A||)^{-1}
,\eeq
which is finite for $\alpha \in (0,1)$.
First we show that $||A^\alpha-B^\beta||$ can be made 
arbitrarily small by choosing $||[\epsilon, A-B]_+||+||[\epsilon,A-B]||_{\ids}$
and $|\alpha-\beta|$ sufficiently small.
\bea
  ||A^\alpha\!\!\!&-&\!\!\!\!B^\beta||\leq{|\sin \alpha \pi |\over \pi}
\int_{0}^{\infty}d\lambda\lambda^{\alpha}4||(\lambda I+B)^{-1}||^2||A-B||\cr\nn
 &+&\!\!\!  |{\sin \alpha \pi \over \pi}- {\sin \beta \pi \over \pi}|
\int_{0}^{\infty}d\lambda\lambda^{\alpha-1}||B||(\lambda+||B||)^{-1}
+ {|\sin \beta \pi |\over \pi}\Big|\int_{0}^{\infty}(\lambda^{\alpha-1}-
\lambda^{\beta-1})||B||(\lambda+||B||)^{-1}
\eea
One can see that all the terms above can be made as small as we wish.
This shows that the diagonal elements satisfy the required continuity.
For the off-diagonal parts;
\bea
||[\!\!\!&\epsilon&\!\!\!\!, A^\alpha-B^\beta]||_{\ids}\leq
 {|\sin \alpha \pi |\over \pi}\int_{0}^{\infty}d\lambda\lambda^\alpha
||(\lambda I+B)^{-1}||^2\Big(||[\epsilon,A-B]||_{\ids}+4||A-B||
||[\epsilon,B]||_{\ids}\Big)\cr\nn
&+&\!\!\!|{\sin \alpha \pi \over \pi}- {\sin \beta \pi \over \pi}|
\int_{0}^{\infty}d\lambda\lambda^{\alpha-1}||(\lambda I+B)^{-1}||\
||[\epsilon,B]||_{\ids}(||B||+||(\lambda I+B)^{-1}||)\cr\nn
&+&\!\!\!{|\sin \alpha \pi |\over \pi}\Big|\int_{0}^{\infty}d\lambda
(\lambda^\alpha-\lambda^\beta)||(\lambda I+B)^{-1}||^2||[\epsilon,B]||_{\ids}
\Big|
,\eea
where all the terms can be made arbitrarily small by choosing a small 
enough neighborhood.
Thus we prove the joint continuity.

Incidentally we remark that the proof does not depend on the specific
ideal.

\end{document}